\documentclass[a4paper,12pt]{article}

\ifx\pdfoutput\undefined
\usepackage[dvips,bookmarks=false]{hyperref}	
\else
\usepackage{hyperref}	
\fi
\hypersetup{colorlinks,bookmarksopen,bookmarksnumbered,citecolor=blue,
linkcolor=black,pdfstartview=FitH,urlcolor=blue}

\usepackage{graphicx}
\usepackage{amsmath}
\usepackage{amssymb}
\usepackage{cite}
\usepackage{braket}
\usepackage{bm}
\usepackage{color}
\usepackage{hhline}
\usepackage{multirow}
\usepackage{indentfirst}
\usepackage{latexsym}

\allowdisplaybreaks

\newcommand{\lra}{\leftrightarrow}

\newcommand{\mc}{\mathcal}
\newcommand{\mr}{\mathrm}

\newcommand{\mbb}{\mathbb}

\numberwithin{equation}{section}

\begin{document}

\begin{titlepage}

\begin{flushright}
KUNS-2849
\end{flushright}

\begin{center}

\vspace{1cm}
{\large\textbf{
Non-thermal Production of PNGB Dark Matter and Inflation
}
 }
\vspace{1cm}

\renewcommand{\thefootnote}{\fnsymbol{footnote}}
Yoshihiko Abe$^{1}$\footnote[1]{y.abe@gauge.scphys.kyoto-u.ac.jp}
,
Takashi Toma$^{2}$\footnote[2]{toma@staff.kanazawa-u.ac.jp}
,
Koichi Yoshioka$^{1}$\footnote[3]{yoshioka@gauge.scphys.kyoto-u.ac.jp}
\vspace{5mm}

\textit{
$^1${Department of Physics, Kyoto University, Kyoto 606-8502, Japan}\\
$^2${Institute of Liberal Arts and Science,\\
Kanazawa University, Kakuma-machi, Kanazawa, 920-1192 Japan}
}

\vspace{8mm}

\abstract{
A pseudo Nambu-Goldstone boson (pNGB)
is a natural candidate of dark matter in that
it avoids the severe direct detection bounds.
We show in this paper that the pNGB has another different and interesting face
with a higher symmetry breaking scale.
Such large symmetry breaking is motivated by various physics beyond
the standard model. In this case, the pNGB interaction is suppressed
due to the Nambu-Goldstone property and the freeze-out production does
not work even with sufficiently large portal coupling.
We then study the pNGB dark matter relic abundance from the
out-of-equilibrium production via feeble Higgs portal coupling.
Further, a possibility is pursued the symmetry breaking scalar in the pNGB model
plays the role of inflaton.
The inflaton and dark matter are unified in a single field
and the pNGB production from inflaton decay is inevitable.
For these non-thermally produced relic abundance of pNGB dark matter
and successful inflation,
we find that the dark matter mass should be less than a few GeV
in the wide range of the reheating temperature and the inflaton mass.
}

\end{center}
\end{titlepage}

\renewcommand{\thefootnote}{\arabic{footnote}}
\newcommand{\bhline}[1]{\noalign{\hrule height #1}}
\newcommand{\bvline}[1]{\vrule width #1}

\setcounter{footnote}{0}

\setcounter{page}{1}

\tableofcontents

\bigskip

\section{Introduction}
\label{sec:intro}

Revealing the evolution of the universe is a key subject not only for
cosmology but also for particle physics. 
In particular, in the situation that there is no clear signature of physics beyond the Standard Model (SM) at the Large Hadron Collider so far, 
exploring nature of dark matter which is presumed to exist in the universe from various observations can give substantial hints for physics beyond the SM.

One of the well-motivated dark matter candidates is so-called WIMPs (Weakly Interacting Massive Particles) which are thermally produced in the early universe 
via sufficient interactions with the SM particles.
WIMPs are widely being searched by various experiments through indirect detection, direct detection, particle collider experiments and astrophysical observations. 
However there is no clear evidence for WIMPs so far, and the resultant experimental constraints become stronger and stronger. 
Direct detection experiments especially give strong constraints on interactions between dark matter and nuclei. 
The current upper bound in terms of WIMP-nucleon spin-independent
cross section is $4.1\times10^{-47}~\mathrm{cm}^2$ at
$30~\mathrm{GeV}$ WIMP mass, which is given by the XENON1T
Collaboration~\cite{Aprile:2018dbl}. 
In addition, as future sensitivity, the XENONnT experiment is expected to update the bound to $2.6\times10^{-48}~\mathrm{cm}^2$ at $50~\mathrm{GeV}$ WIMP mass~\cite{Aprile:2020vtw}.
Such a severe constraint may imply that the interactions between dark
matter and SM particles are rather weak, which means less motivation for thermal WIMP paradigm.

One of the ways naturally evading the severe constraints from direct
detection is to identify a pseudo Nambu-Goldstone boson (pNGB) as dark matter.
The simplest model of pNGB dark matter with global $U(1)$ symmetry
breaking has been proposed in Ref.~\cite{Gross:2017dan} and 
its extensions with gauged $U(1)_{B-L}$ symmetry are discussed in
Refs.~\cite{Abe:2020iph, Okada:2020zxo}.
In addition, various aspects of the pNGB dark matter model, such as a global fit \cite{Arina:2019tib}, have been studied.
In these cases, it is found due to the nature of NGB that all
couplings of dark matter are inversely proportional to the vacuum
expectation value (VEV) associated with the symmetry breaking, and
then highly suppressed if the VEV is large enough.

Such a large VEV may in fact be connected to generate the small neutrino masses in the framework of Majoron models where the right-handed neutrino Majorana masses are induced by the large VEV~\cite{Chikashige:1980qk,Chikashige:1980ui,Gelmini:1980re,Gu:2010ys,Matsumoto:2010hz,Queiroz:2014yna}. 
In this case, the pNGB is identified as Majoron. 
In order to make the canonical seesaw mechanism work with $\mathcal{O}(1)$
Yukawa couplings, the VEV should be as large 
as $\mathcal{O}(10^{14})~\mathrm{GeV}$. 
Therefore from this viewpoint, it is motivated to consider the pNGB as
FIMPs (Feebly Interacting Massive Particles) produced by freeze-in 
mechanism~\cite{Asaka:2005cn, Asaka:2006fs, Hall:2009bx} with extremely suppressed interactions, e.g., due to a large VEV\@.
In the framework of freeze-in mechanism, dark matter is assumed to be never thermalized with the SM particles. 
A typical magnitude of FIMP coupling for reproducing the relic abundance observed by the PLANCK Collaboration~\cite{Aghanim:2018eyx} is 
$\mathcal{O}(10^{-11})$ for dimensionless couplings~\cite{Hall:2009bx}.

In this paper, we calculate the dark matter relic abundance via Higgs
portal in the pNGB dark matter model~\cite{Gross:2017dan} with large
symmetry breaking scale,
and study in detail the freeze-in parameter space consistent with the observations. 
In addition, we examine if successful inflation can occur through
the non-minimal coupling to gravity where the field associated with
the symmetry breaking is identified as the inflaton.
That implies the inflaton also necessarily induces the pNGB dark
matter relic, which would significantly modify the relevant parameter space.

The rest part of this paper is organized as follows.
In Sec.~\ref{sec:model}, we briefly review the pNGB model.
In Sec.~\ref{sec:FI},
the relic abundance of pNGB dark matter via the Higgs portal freeze-in
is calculated, including the thermal mass of the Higgs boson which is important to evaluate the reaction rates of relevant processes. 
We also derive the Boltzmann equations for the pNGB FIMP,
evaluate the time evolution of the dark matter yield, and show
some parameter sets consistent with the present relic abundance
observed by experiments. 
In Sec.~\ref{sec:inflation}, we examine the possibility 
that the radial component of symmetry breaking scalar plays an role of the inflaton.
The allowed parameter space is identified taking into account
the direct production of the pNGB dark matter from the inflaton
decay. Sec.~\ref{sec:conclusion} is devoted to our conclusion.

\section{PNGB Dark Matter Model}
\label{sec:model}

In the pNGB model, the SM is extended with a complex singlet scalar $\Phi$, and the Lagrangian is given by
\begin{align}
 \mc{L} = \mc{L}_{\mathrm{SM}} + |\partial_\mu \Phi|^2 - \mc{V}(H, \Phi),
\end{align}
where the scalar potential including the SM Higgs doublet $H$ is given by
\begin{align}
 \mc{V} (H, \Phi) = & - \frac{\mu_H^2}{2} |H|^2 + \frac{\lambda_H}{2} |H|^4
 - \frac{\mu_\Phi^2}{2} |\Phi|^2 + \frac{\lambda_\Phi}{2} |\Phi|^4 + \lambda_{H\Phi} |H|^2 |\Phi|^2
 \nonumber\\
 & - \frac{m^2}{4} \bigl( \Phi^2 + \Phi^{*2} \bigr).
\end{align}
The last term is the soft-breaking mass term which is introduced in
order to generate the mass of pNGB\@.
We do not consider the origin of
this term (ultraviolet (UV) 
completion of the model), while some examples have been discussed in
the literature~\cite{Abe:2020iph, Okada:2020zxo}. 
The Higgs doublet $H$ and the singlet scalar $\Phi$ are assumed to
develop non-vanishing VEVs and are parametrized as
\begin{align}
 H = \frac{1}{\sqrt{2}} \left( \begin{array}{c}
 0 \\
 v + h
 \end{array}\right),
 \qquad
 \Phi = \frac{v_\phi + \phi}{\sqrt{2}}\,e^{i \chi / v_\phi},
 \label{eq:fluctuation}
\end{align}
where we have dropped the would-be NG modes in $H$ (the unitary gauge).
Note that the pNGB $\chi$ is stable due to a remnant $\mathbb{Z}_2$ symmetry after the spontaneous symmetry breaking, thus it can be a dark matter candidate.
The stationary conditions of the VEVs $v$ and $v_\phi$ impose the following relations between the parameters in the scalar potential
\begin{align}
 \mu_H^2 = \lambda_H v^2 + \lambda_{H\Phi} v_\phi^2,
 \qquad
 \mu_\Phi^2 = \lambda_\Phi v_\phi^2 + \lambda_{H\Phi} v^2 -m^2.
\end{align}
Using these relations, the masses of the scalar fields are evaluated
in the following two phases:
\begin{itemize}
\item
The electroweak symmetry is unbroken,
$\Braket{H}=0$ and $\braket{\Phi} = v_\phi /\sqrt{2}$.
In this phase, only the components of $\Phi$ acquire the masses as
\begin{align}
 m_\phi^2 = \lambda_\Phi v_\phi^2 ,
 \qquad
 m_\chi^2 = m^2.
\end{align}

\item
The electroweak symmetry is spontaneously broken,
$\Braket{H} = ( 0, v/\sqrt{2} )^{\mr{T}}$ and $\braket{\Phi} = v_\phi /\sqrt{2}$.
In this phase, the physical component of $H$ is massive and the mass
eigenvalues of the scalar fields are given by 
\begin{align}
 m_{h_1}^2 = &~ \frac{1}{2} \biggl[ \lambda_H v^2 + \lambda_\Phi v_\phi^2 
 - \sqrt{ ( \lambda_\Phi v_\phi^2 - \lambda_H v^2 )^2 + 4 \lambda_{H\Phi}^2 v^2 v_\phi^2} \biggr],
 \\
 m_{h_2}^2 = &~ \frac{1}{2} \bigg[ \lambda_H v^2 + \lambda_{\Phi} v_\phi^2 +\sqrt{ ( \lambda_\Phi v_\phi^2 - \lambda_H v^2 )^2 + 4 \lambda_{H\Phi}^2 v^2 v_\phi^2}\biggr],
 \\[1mm]
 m_{\chi}^2 =&~ m^2.
\end{align}
The lighter CP-even scalar $h_1$ is identified as the SM-like Higgs
boson with the mass 125 GeV\@.
The mixing angle of the CP-even scalars is given by
\begin{align}
 \tan 2 \alpha = \frac{2 \lambda_{H\Phi} v v_\phi}{\lambda_\Phi v_\phi^2-\lambda_H v^2},
\end{align}
which is introduced as
\begin{align}
 \left( \begin{array}{c}
 h \\
 \phi
 \end{array}\right)
 = \left( \begin{array}{cc}
 \cos \alpha & \sin \alpha\\
 - \sin \alpha & \cos \alpha
 \end{array}\right)
 \left( \begin{array}{c}
 h_1 \\
 h_2
 \end{array} \right).
\end{align}
Note that the mixing can safely be ignored in our
setup with a large hierarchy between the VEVs ($v\ll v_{\phi}$).
\end{itemize}

In this paper, we mainly use the non-linear representation for the
fluctuations of $\Phi$ field as given in Eq.~\eqref{eq:fluctuation}.
The same physics is obtained also in the linear representation.
The Lagrangian in the broken phase of $\Phi$ contains
\begin{align}
 \mc{L} \supset &~ 
 \frac{1}{2} \Bigl[ (\partial_\mu \phi)^2 - m_\phi^2 \phi^2 \Bigr]
 + \frac{1}{2} \biggl( 1 + \frac{\phi}{v_\phi} \biggr)^2 \biggl[
 (\partial_\mu \chi )^2 - m_\chi^2 v_\phi^2 \sin^2 \biggl( \frac{\chi}{v_\phi} \biggr) \biggr]
 \nonumber\\
 &~ -\frac{\lambda_\Phi}{8} \phi^4 - \frac{\lambda_\Phi v_\phi}{2} \phi^3
 -\frac{\lambda_{H\Phi}}{2} |H|^2 \bigl( v_\phi + \phi \bigr)^2.
\end{align}
The interaction terms of pNGB dark matter $\chi$ are originated from
the scalar kinetic term and the $U(1)$ soft-breaking mass
term in this representation.

\section{PNGB Production via Freeze-in}
\label{sec:FI}

In this section,  
we discuss how the pNGB dark matter relics are produced via the 
freeze-in mechanism and the parameter set consistent with the
observations. We assume that the relic abundance of dark matter is
determined in the radiation dominant era of the universe, 
in which the Hubble parameter $H$ and entropy density $s$ are given as
the functions of the temperature $T$ as
\begin{align}
 H = \sqrt{ \frac{\pi^2}{90} g_* }\frac{T^2}{M_P},
 \qquad
 s = \frac{2\pi^2}{45} g_*^S T^3,
\end{align}
where $g_*$ and $g_*^S$ denote the total numbers of effective massless
degrees of freedom contributing to the energy and entropy densities, respectively~\cite{Kolb:1990vq}, and $M_P$ is the reduced Planck mass $M_P = 1/\sqrt{8\pi G_N} = 2.4 \times 10^{18}~\mr{GeV}$ 
($G_N$ is the gravitational constant).

\subsection{Boltzmann equations}

We are interested in the case that both of dark matter $\chi$ and the CP-even
scalar $\phi$ ($\simeq h_2$) are never thermalized with the SM particles. 
This is achieved by tiny values of quartic couplings $\lambda_{H\Phi}$ and
$\lambda_{\Phi}$, roughly speaking, 
$\lambda_{H\Phi},\lambda_{\Phi}\lesssim10^{-6}$. From the theoretical
side, a radiative correction to the $|\Phi|^4$ term would imply its
lower bound, $\lambda_\Phi\gtrsim \lambda_{H\Phi}^2/16 \pi^2$.
With these feeble couplings, both of $\phi$ and
$\chi$ become the FIMPs, and the Boltzmann equations for the number
densities $n_\phi$ and $n_\chi$ are given by
\begin{align}
 \frac{d n_\phi}{dt} + 3H n_\phi =&~ C_{H^\dagger H \lra \phi \phi} + C_{\chi \chi \lra \phi \phi} + C_{H^\dagger H \lra \phi} + C_{\chi \chi \lra \phi},
 \\
 \frac{d n_{\chi}}{dt} + 3 H n_\chi =&~ C_{\phi \phi \lra \chi \chi} + C_{H^\dagger H \lra \chi \chi} + C_{\phi \lra \chi \chi},
\end{align}
where $C_{A \lra B}$ in the right-hand side denotes the collision term
corresponding to the process $A \lra B$. 
Here the broken phase of $\Phi$ is assumed and the scalars in the dark sector 
interact with the SM only through the Higgs doublet. The
explicit form of collision terms is 
\begin{align}
 C_{ij\ldots \lra\,ab\ldots} = & 
 \int \prod_i d\Pi_i f_i \prod_a d\Pi_a (1+f_a) \; (2\pi)^4 
 \delta^4 \biggl(\sum_ip_i - \sum_a p_a \biggr)\, 
 |\mc{M}_{ij\ldots \to\,ab\ldots}|^2
 \nonumber \\
 & -\int \prod_a d\Pi_a f_a \prod_i d\Pi_i (1+f_i) \; (2\pi)^4 
 \delta^4 \biggl(\sum_ap_a - \sum_i p_i \biggr)\, 
 |\mc{M}_{ab\ldots \to\,ij\ldots}|^2,
\end{align}
where $f_x$ is the distribution function of particle $x$,
$d\Pi_x$ is the Lorentz-invariant phase space expressed 
as $ d\Pi_x = \frac{d^3 \bm{p}_x}{(2\pi)^3 2 E_{\bm{p}_x}}$,
and $\mc{M}_X$ denotes the amplitude of the process $X$.

Since dark matter is produced by the freeze-in mechanism from the SM thermal bath,
the magnitude of distribution functions is tiny for $\phi$ and $\chi$. 
Thus the $\mc{O}(f_{\phi,\chi}^2)$ terms can be dropped in the above equations.
This approximation is valid as long as the distribution functions 
$f_{\phi, \chi}$ are not close to 
the equilibrium one before the abundance is frozen.
The Boltzmann equations are thus reduced to 
\begin{align}
 \frac{d n_{\phi}}{dt} + 3 H n_\phi = &~ C_{H^\dagger H \to \phi \phi} + C_{H^\dagger H \to \phi}
 - \int d\Pi_\phi\, f_\phi(\bm{p}_\phi)\, 2 m_{\phi} 
 \bigl( \Gamma_{\phi \to H^\dagger H} + \Gamma_{\phi \to \chi \chi} \bigr),
 \\
 \frac{d n_{\chi}}{dt} + 3 H n_{\chi} = &~ C_{H^\dagger H \to \chi \chi}
 + 2 \int d\Pi_\phi\, f_\phi(\bm{p}_\phi)\, 2 m_\phi \Gamma_{\phi \to \chi \chi},
\end{align}
where the decay widths are given by
\begin{align}
 \Gamma_{\phi \to H^\dagger H} = \frac{\lambda_{H\Phi}^2 m_\phi}{8 \pi \lambda_\Phi} \sqrt{ 1 - \frac{4 m_H^2}{m_\phi^2}},
 \qquad
 \Gamma_{\phi \to \chi \chi} = \frac{\lambda_\Phi m_\phi}{32 \pi} \sqrt{ 1 - \frac{4 m_\chi^2}{m_\phi^2}}.
\end{align}
Note that we have used the relation $m_{\phi}^2=\lambda_{\Phi}v_{\phi}^2$ in the above equations and the SM Higgs doublet contains real four components, leading to
the difference of numerical factors. The mass parameter
$m_H$ for the $H$ field will be discussed in the next section. 
Introducing the net dark matter number density $n_D = n_\chi +2\mr{Br}^{\phi \to \chi \chi} n_\phi$,
the Boltzmann equation for $n_D$ is recast as
\begin{align}
 \frac{d n_D}{dt} + 3 H n_D = C_{H^\dagger H \to \chi \chi} + 2 \mr{Br}^{\phi \to \chi \chi} \bigl( C_{H^\dagger H \to \phi \phi} + C_{H^\dagger H \to \phi} \bigr),
 \label{eq:boltzmanneqnD}
\end{align}
where $\mr{Br}^{\phi \to \chi \chi}$ is the branching ratio defined by $\mr{Br}^{\phi \to \chi \chi} = \Gamma_{\phi \to \chi \chi} / ( \Gamma_{\phi \to H^\dagger H} + \Gamma_{\phi \to \chi \chi})$.
The collision terms are written by using the thermally
averaged cross sections and the number density of the SM Higgs doublet
in thermal bath, $n_H^{\mr{eq}}$. Then we obtain 
\begin{align}
 \frac{d n_{D}}{dt} + 3 H n_D = 
   2 \left[\braket{ \sigma_{H^\dagger H \to \chi \chi} \bar{v}}
 + 2 \mr{Br}^{\phi \to \chi \chi}
 \braket{ \sigma_{H^\dagger H \to \phi \phi} \bar{v} }
 + \mr{Br}^{\phi \to \chi \chi}
 \braket{ \sigma_{H^\dagger H \to \phi} \bar{v} }\right] (n_H^{\mr{eq}})^2.
 \label{eq:boltzmanneqnD2}
\end{align}
The first term in the right-hand side denotes the dark matter
production directly from the thermal bath,
and the second and third terms are the contributions from the
decays of $\phi$ produced from the thermal bath.
The thermally averaged cross sections are explicitly
calculated as
\begin{align}
 \braket{\sigma_{H^\dagger H \to \chi \chi} \bar{v}} (n^{\mr{eq}}_H)^2 =&~ \frac{\lambda_{H\Phi}^2T^4}{256 \pi^5}
 \int_{2 \bar{x}_\chi}^\infty dz\, \sqrt{z^2 - 4 x_\chi^2} \sqrt{z^2 - 4 x_H^2} \frac{z^4K_1( z )}{(z^2 - x_\phi^2)^2 + x_\phi^2 \gamma_\phi^2},
 \label{eq:RRHHchichi}
 \\
 \braket{\sigma_{H^\dagger H \to \phi \phi} \bar{v}} (n^{\mr{eq}}_H)^2 =&~ \frac{\lambda_{H\Phi}^2T^4}{256 \pi^5}
 \int_{2 \bar{x}_\phi}^\infty dz\, \sqrt{z^2 - 4x_\phi^2} \sqrt{z^2 - 4 x_H^2} 
 \frac{( z^2 + 2 x_\phi^2)^2 + x_\phi^2 \gamma_\phi^2}{(z^2 - x_\phi^2)^2 + x_\phi^2 \gamma_\phi^2} K_1 (z),
 \label{eq:RRHHphiphi}
 \\
 \braket{\sigma_{H^\dagger H \to \phi} \bar{v} } (n_H^{\mr{eq}})^2 =&~ \frac{\lambda_{H\Phi}^2 m_\phi^3 T}{16 \pi^3\lambda_\Phi} K_1 (x_\phi) \sqrt{1 - \frac{4m_H^2}{m_\phi^2} },
 \label{eq:RRHHphi}
\end{align}
where we have defined the integral variable $z\equiv\sqrt{s}/T$
for the Mandelstam $s$ variable, 
the dimensionless parameters $x_i \equiv m_i /T$ 
and $\gamma_i \equiv \Gamma_i / T$ for a particle $i$ with the mass
$m_i$ and the total decay width $\Gamma_i$. 
In the integrals, the lower limits are given 
by $\bar{x}_{\chi,\phi}=\max[x_{\chi,\phi},x_H]$ and $K_1(z)$ is the
modified Bessel function of second kind of order 1. 
Here thermal bath particles are assumed to obey the Maxwell-Boltzmann
distribution.\footnote{
Quantum statistical distributions may give a small factor difference in numerical calculations~\cite{Lebedev:2019ton}.
}
\ 
Note that the integrand in Eq.~(\ref{eq:RRHHchichi}) contains the
factor $z^4$ due to the derivative coupling of pNGB dark matter.
This behavior implies that even if the portal coupling $\lambda_{H\Phi}$
is large, the dark matter reaction rate with the SM is suppressed in
lower energy than $m_\phi$ and the usual freeze-out does not work
in the case of the hierarchical VEV $v_\phi \gg v\,$ ($m_\phi \gg m_h$).

\subsection{Thermal mass of the Higgs boson}

The one-loop effect of bath particles leads to the mass corrections
quadratically scaling by the temperature,
which are called thermal masses.
Including thermal mass corrections gives important effects for
collision terms (reaction rates) and hence for the evaluation of relic
abundance of dark matter. 
The detailed calculation of thermal mass for the SM Higgs boson is
summarized in Appendix~\ref{sec:Tmass}.
The electroweak gauge bosons, all SM fermions, and
all components of $H$ including the NGBs eaten by the gauge bosons
contribute to the thermal mass, which is given by
\begin{align}
 \Delta_h = \biggl( \frac{g_Y^2}{16} + \frac{3g_2^2}{16} + \frac{y_t^2}{4} + \frac{\lambda_H}{4} \biggr) T^2,
\end{align}
where $g_Y$ and $g_2$ are the gauge couplings of $U(1)_Y$ and
$SU(2)_L$, and $y_t$ is the top Yukawa coupling which is dominant over
the other Yukawa couplings. 
Including this thermal contribution, the mass of the SM Higgs boson is given by
\begin{align}
 m_H^2 = m_{h_1}^2 + \Delta_h,
\end{align}
which plays a role of regulator in the reaction rates in the Boltzmann equations.

\subsection{IR freeze-in ~$(T_R\gg m_\phi)$}
\label{sec:IRFI}

When the reheating temperature of the universe, $T_R$, is higher than the heavy mediator mass $m_{\phi}$, the final abundance of dark matter is determined independently of the reheating temperature.
This is so-called the IR freeze-in~\cite{Hall:2009bx}. 
The analytic formula of the dark matter yield $Y_{D}=n_{D}/s$ derived from Eq.~\eqref{eq:boltzmanneqnD2} is approximately given by
\begin{align}
 Y^{\mr{IR}}_D \approx
 \frac{405 \sqrt{10}}{\left(2\pi\right)^5}\,
 \frac{\mr{Br}^{\phi\to \chi \chi}\lambda_{H\Phi}^2 M_P}{g^S_* g_*^{1/2} \lambda_\Phi m_\phi}
 \sqrt{ 1 - \frac{4m_h^2}{m_\phi^2}}.
 \label{eq:relicapprox}
\end{align}
The result is found to be independent of the dark matter mass
because the dominant production process comes from the $\phi$ decay
around $T \sim m_\phi$, as shown in the following part.
In the limit where the scalar VEV and the mass of $\phi$
are much larger than the masses of the SM Higgs boson and dark matter,
the factor of quartic couplings $\mr{Br}^{\phi \to \chi \chi} \lambda_{H\Phi}^2$ reduces to
\begin{align}
 \mr{Br}^{\phi \to \chi \chi} \lambda_{H\Phi}^2 \approx
 \frac{\lambda_{\Phi}^2 \lambda_{H\Phi}^2}{\lambda_{\Phi}^2 + 4 \lambda_{H\Phi}^2}
 \approx
 \begin{cases}
 \lambda_{H\Phi}^2 & \text{for }\;\lambda_{\Phi} \gg \lambda_{H\Phi}\\
 \frac{\lambda_{\Phi}^2}{4} & \text{for }\;\lambda_{\Phi} \ll \lambda_{H\Phi}
 \end{cases}.
\end{align}
Thus combining the IR freeze-in relic Eq.~\eqref{eq:relicapprox} and the yield corresponding to the observed value $Y_D^\mathrm{obs.}=4.4\times10^{-7}\left(m_{\chi}/\rm{MeV}\right)^{-1}$~\cite{Aghanim:2018eyx} 
with $g_*\sim g_*^S\sim100$, we obtain the following relations
\begin{align}
 \left(\frac{\lambda_{\Phi}}{10^{-8}}\right)
 &\approx \left(\frac{\lambda_{H\Phi}}{2\times10^{-9}}\right)^2
 \biggl(\frac{m_{\chi}}{1~\mr{MeV}}\biggr)
 \left(\frac{10^{10}~\rm{GeV}}{m_{\phi}}\right)
 &\text{for }\;\lambda_{\Phi}\gg \lambda_{H\Phi},
 \label{eq:ir_hphi}\\
 \left(\frac{\lambda_{\Phi}}{7\times10^{-11}}\right)
 &\approx \left(\frac{1~\mr{MeV}}{m_{\chi}}\right)
 \biggl(\frac{m_{\phi}}{10^{10}~\rm{GeV}}\biggr)
 &\text{for }\;\lambda_{\Phi}\ll \lambda_{H\Phi}.
 \label{eq:ir_phi}
\end{align}

\begin{figure}[t]
\centering
\includegraphics[scale=0.63]{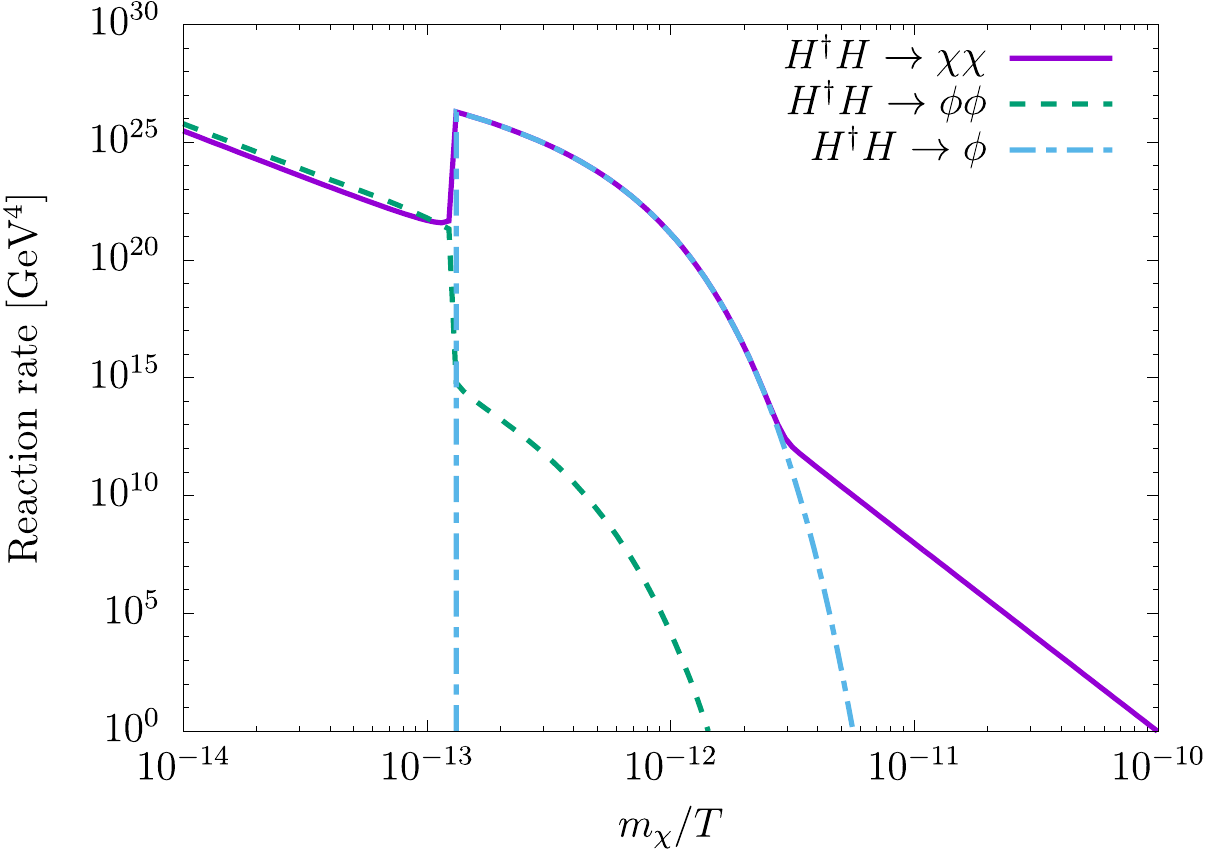} ~
\includegraphics[scale=0.63]{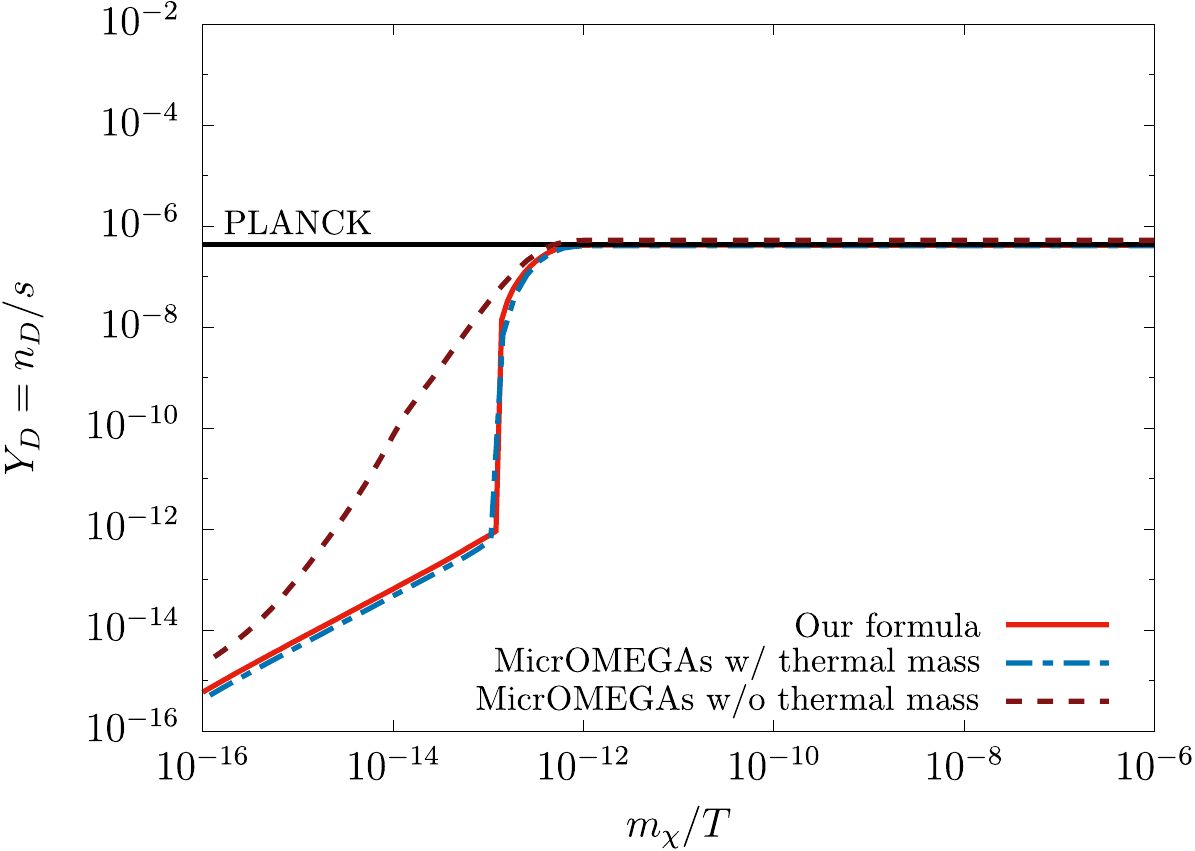}
\caption{
(Left):
Evolution of the reaction rates for three processes
Eqs.~\eqref{eq:RRHHchichi}--\eqref{eq:RRHHphi}.
(Right):
Evolution of the dark matter yield determined by solving the
Boltzmann equation. 
The red solid, dot-dashed dark blue and dashed brown lines 
denote the results using our formula~\eqref{eq:boltzmanneqnD2}, the
public code MicrOMEGAs with and without the thermal mass effect for
the SM Higgs boson, respectively. 
The black horizontal line corresponds to the correct dark matter
abundance for $m_\chi =1~\mr{MeV}$.
\bigskip}
\label{fig:tdep}
\end{figure}

The evolution of the reaction rates including the thermal mass effects
is shown in the left panel of Fig.~\ref{fig:tdep} where we choose the
following parameter set 
\begin{align}
 \lambda_\phi = 7\times10^{-11},
 \quad
 \lambda_{H\Phi} = 10^{-7},
 \quad
 m_\chi = 1~\mr{MeV},
 \quad
 m_\phi = 10^{10}~\mr{GeV},
 \label{eq:bench}
\end{align}
as a benchmark. This is chosen so that
the parameter relation Eq.~\eqref{eq:ir_phi} is realized. 
Note that the portal coupling should satisfy $\lambda_{H\Phi} \lesssim 10^{-6}$
in order for the dark sector particles not to enter into the thermal bath.
When the temperature cools down to the mediator mass 
scale $T\sim m_\phi$, a number of on-shell $\phi$ are resonantly produced. 
As a result the magnitude of the reaction rate for the 
process $H^{\dag}H\to \chi\chi$ rapidly increases as can be seen in Fig.~\ref{fig:tdep}.

For the same benchmark parameter set, the evolution of the dark matter
yield is shown in the right panel of Fig.~\ref{fig:tdep}, assuming
the vanishing initial conditions of the dark sector
\begin{align}
 Y_{\chi}(T= T_R) = Y_{\phi}(T = T_R) =0.
\end{align}
We also show for comparison the results calculated by the public code
MicrOMEGAs~\cite{Belanger:2018ccd} with and without the thermal mass effect for the SM Higgs boson.\footnote{
In the current version of MicrOMEGAs, full thermal mass effects are not implemented. However it is possible to include the thermal mass only for the Higgs boson by hand without difficulty.
}
\ 
As obvious from the plot, the thermal mass effect gives an impact on
the evolution of the dark matter yield, in particular when the
temperature is $T \gtrsim m_{\phi}$.
Including thermal mass also affects the final dark matter abundance,
while its impact is not so large and only gives a few factor difference. 
One can see from the plot that the yield of dark matter rapidly grows
at $T \sim m_\phi$ due to the resonant production of $\phi$, and then
the evolution is almost frozen afterwards.

\begin{figure}[t]
\centering
\includegraphics[scale=0.63]{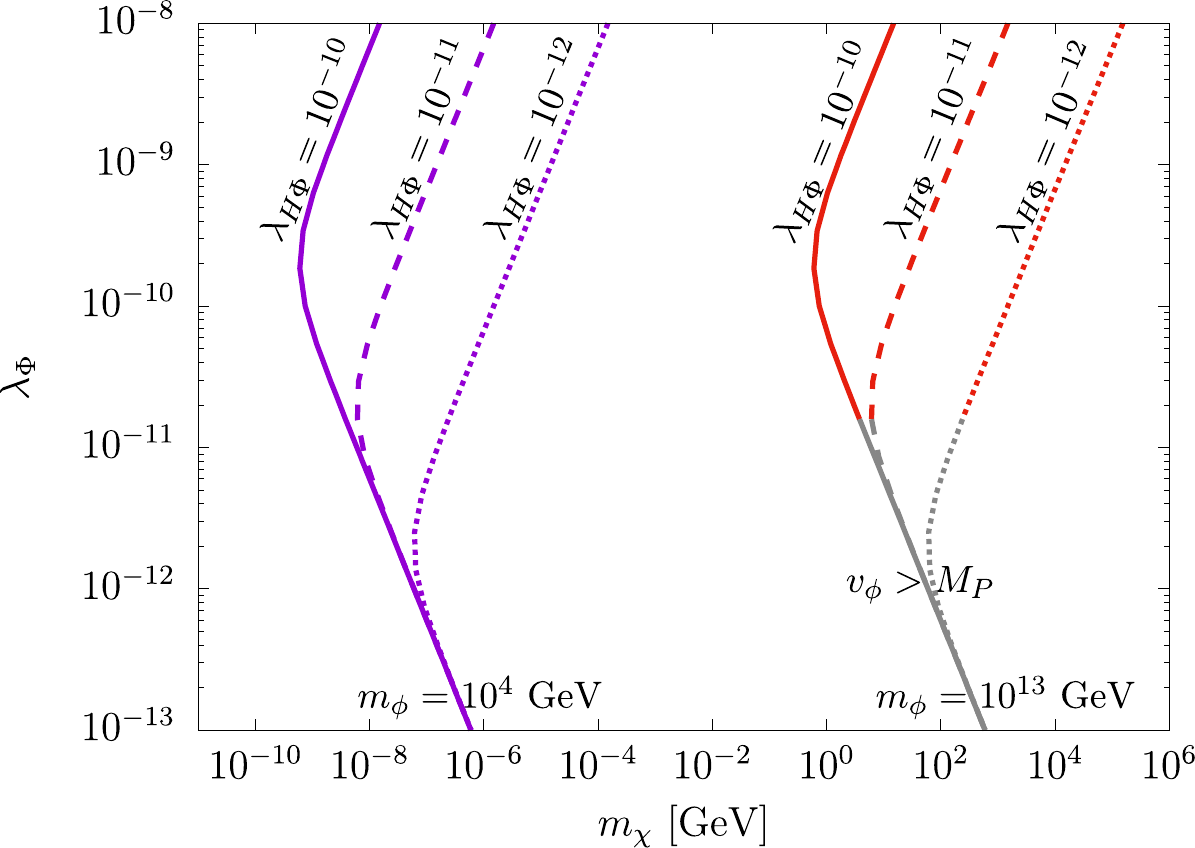} ~~
\includegraphics[scale=0.41]{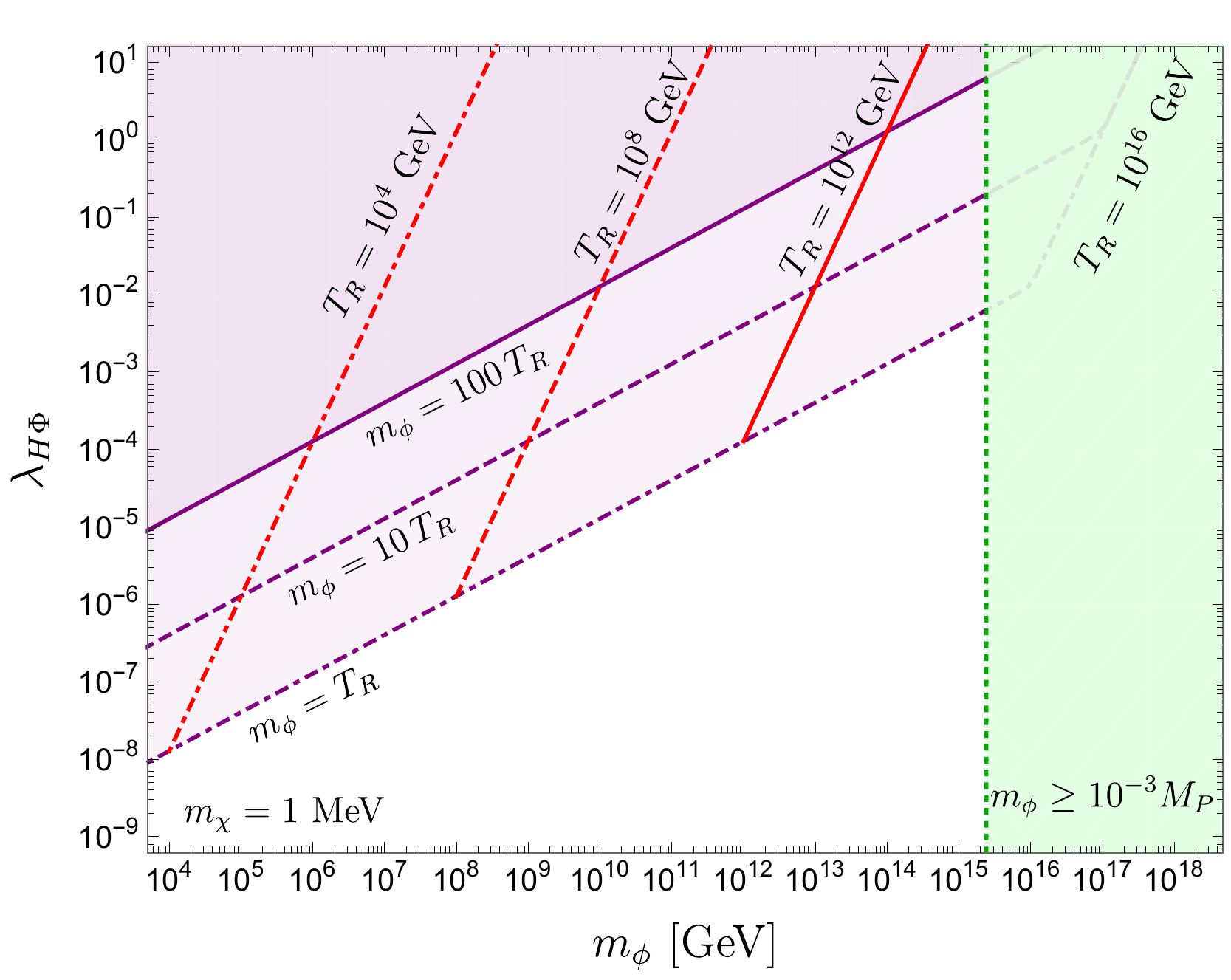}
\caption{
(Left):
Contours reproducing the observed relic abundance in 
the $(m_\chi , \lambda_{\Phi})$ plane where the mediator mass is fixed to be
$m_{\phi}=10^{4}~\mathrm{GeV}$ ($10^{13}~\mathrm{GeV}$) in the
purple (red) line. The solid, dashed and dotted lines correspond 
to $\lambda_{H\Phi}=10^{-10}$, $10^{-11}$ and $10^{-12}$, respectively.
The gray parts of the red lines are excluded by the criterion $v_\phi >M_P$.
(Right):
The purple region can realize the observed relic abundance via the UV
freeze-in in the $(m_\phi, \lambda_{H\Phi})$ plane. 
Each contour reproduces the abundance for a specific parameter set. 
The solid, dashed and dot-dashed red lines correspond 
to $T_R = 10^{12}~\mr{GeV}$, $10^{8}~\mr{GeV}$ and $10^{4}~\mr{GeV}$, respectively.
The green region, $m_\phi \geq 10^{-3} M_P$, is excluded (see the text).
\bigskip}
\label{fig:allowedregion}
\end{figure}

The purple and red lines in the left panel of
Fig.~\ref{fig:allowedregion} show the 
contours in the ($m_{\chi}$, $\lambda_{\Phi}$) plane reproducing the
dark matter relic abundance observed 
by the PLANCK Collaboration~\cite{Aghanim:2018eyx}.
The reheating temperature is assumed to be much higher than the
mediator mass scale ($T_R\gg m_{\phi}$).
As can be seen in the plots, for a larger $m_{\phi}$ the lines
simply shift to the right (purple to red), namely the direction of
heavier dark matter mass which compensates a smaller yield $Y_D^\mr{IR}$.
The behavior of the lines change around $\lambda_{\Phi}\sim\lambda_{H\Phi}$ as explained in Eqs.~(\ref{eq:ir_hphi}) and (\ref{eq:ir_phi}). 
While one can take a heavier dark matter mass $m_\chi$ if a larger mediator
mass $m_{\phi}$ is chosen, 
there is an upper bound on $m_\chi$ if the criterion for
the VEV $v_{\phi}<M_P$ is taken into account. As we will discuss in
the next section, a feeble value of the scalar self-coupling, typically
$\lambda_{\Phi} \gtrsim \mc{O}(10^{-12})$ is suitable for $\phi$ being
the inflaton.

\subsection{UV freeze-in ~$(T_R\ll m_{\phi})$}
\label{sec:UVFI}

When $m_\phi$ is much larger than the reheating temperature,
the dark matter relic abundance is determined by the portal coupling
$\lambda_{H\Phi}$ and the reheating temperature $T_R$.
This is so called the UV freeze-in discussed in
Refs.~\cite{Hall:2009bx,Elahi:2014fsa}. 
For $T<T_R\ll m_{\phi}$, only the $H^{\dag}H\to\chi\chi$ process is effective
for the dark matter production (see the left panel of Fig.~\ref{fig:tdep}).
In the $H^{\dag}H\to\chi\chi$ reaction rate \eqref{eq:RRHHchichi},
we can safely assume $m_\phi \gg \sqrt{s} \gg m_h, m_\chi$ 
since the modified Bessel function regarded as the window function
with a cut-off $T$ and the large $s$ contribution is dominant 
due to the $s^{5/2}$ behavior of the integrand.
Thus the reaction rate is approximated by
\begin{align}
 \braket{\sigma_{H^\dagger H \to \chi \chi} \bar{v}} (n_{H}^{\mr{eq}})^2 \approx \frac{\lambda_{H\Phi}^2T}{512 \pi^5} \int_0^\infty ds\, \frac{s^{5/2}}{m_\phi^4} K_1(\sqrt{s}/T) = \frac{3 \lambda_{H\Phi}^2 T^8}{2\pi^5 m_\phi^4}.
\end{align}
By integrating the Boltzmann equations for the dark sector using this
approximation, the pNGB dark matter yield is evaluated as
\begin{align}
 Y^{\mr{UV}}_D
 \approx \frac{135 \sqrt{10} \lambda_{H\Phi}^2 M_P T_R^3}{4 \pi^{8} g_*^S g_*^{1/2} m_\phi^4}.
 \label{eq:YUV}
\end{align}
Combining with the observed value $Y_D^{\mathrm{obs.}}=4.4\times10^{-7}\left(m_{\chi}/\rm{MeV}\right)^{-1}$ with $g_*\sim g_*^S\sim100$, 
we obtain the following relation\footnote{
For a heavier pNGB dark matter case,
the broader parameter space in the $(m_\phi, \lambda_{H\Phi})$ plane
has been discussed in Ref.~\cite{Abe:2020dut}.}
\begin{align}
\left(\frac{\lambda_{H\Phi}}{10^{-6}}\right)^2\approx
\left(\frac{1~\rm{MeV}}{m_{\chi}}\right)
\biggl(\frac{m_{\phi}}{10^{5}~\rm{GeV}}\biggr)^4
\left(\frac{10^{4}~\rm{GeV}}{T_R}\right)^3.
\label{eq:uv5}
\end{align}

The right panel of Fig.~\ref{fig:allowedregion} shows the parameter
space where the dark matter relic can be realized by the UV freeze-in, 
which is denoted by the purple region.
The purple solid, dashed and dot-dashed lines 
denote $m_{\phi} = 100 \,T_R$, $10\,T_R$ and $T_R$ (the 
UV freeze-in \eqref{eq:YUV} is valid for $m_\phi \gg T_R$).
The red lines show the contours 
in the $(m_\chi, \lambda_{H\Phi})$ plane reproducing
the dark matter relic
for various given reheating temperatures.
Note that the green colored region $m_\phi \geq 10^{-3} M_P$ is
excluded by the conditions $v_\phi \leq M_P$ and 
$\lambda_{\Phi} \leq 10^{-6}$, the latter of which is required for the
dark sector not being thermalized.

\section{Inflation}
\label{sec:inflation}

In the previous section,
we have seen the pNGB dark matter relic abundance realized via feeble
scalar couplings, where the radial component $\phi$ of the symmetry breaking
scalar plays a role of the mediator of pNGB production. In this section,
we investigate the possibility that $\phi$ also plays another
important role, namely, the inflaton.\footnote{
Some different types of
models are studied in the
literature, e.g.~\cite{Boucenna:2014uma,Enqvist:2014zqa} about
possible common origins of the inflaton and dark matter.}

\subsection{Inflationary dynamics and constraints}

The inflation gives a plausible solution for the flatness and horizon
problems in the universe. 
A successful inflation scenario can occur in our model if a non-minimal coupling
between the complex scalar $\Phi$ and gravity is introduced. Then the
Lagrangian relevant for the inflation dynamics is given by
\begin{equation}
 \frac{\mc{L}}{\sqrt{-g}}=-\frac{M_P^2}{2}\mc{R}-\xi|\Phi|^2\mc{R}+g^{\mu\nu}\left(\partial_{\mu}\Phi\right)^*\left(\partial_{\nu}\Phi\right)-\mc{V}(\Phi),
 \label{eq:infLag}
\end{equation}
where $\mc{R}$ is the Ricci scalar 
and $\xi$ is the so-called non-minimal coupling constant. 
During the inflation era, the scalar potential is assumed to be
dominated by the field value of $\Phi$, 
thus $\mc{V}(\Phi) \approx \lambda_{\Phi}|\Phi|^4/2 \approx \lambda_{\Phi}\phi^4/8$,
and the $|\Phi|^2$ part can be written as $|\Phi|^2\approx \phi^2/2$ 
with the non-linear representation $\Phi=\left(v_\phi+\phi\right)e^{i\chi/v_\phi}/\sqrt{2}$.

\begin{figure}[t]
\centering
\includegraphics[scale=0.45]{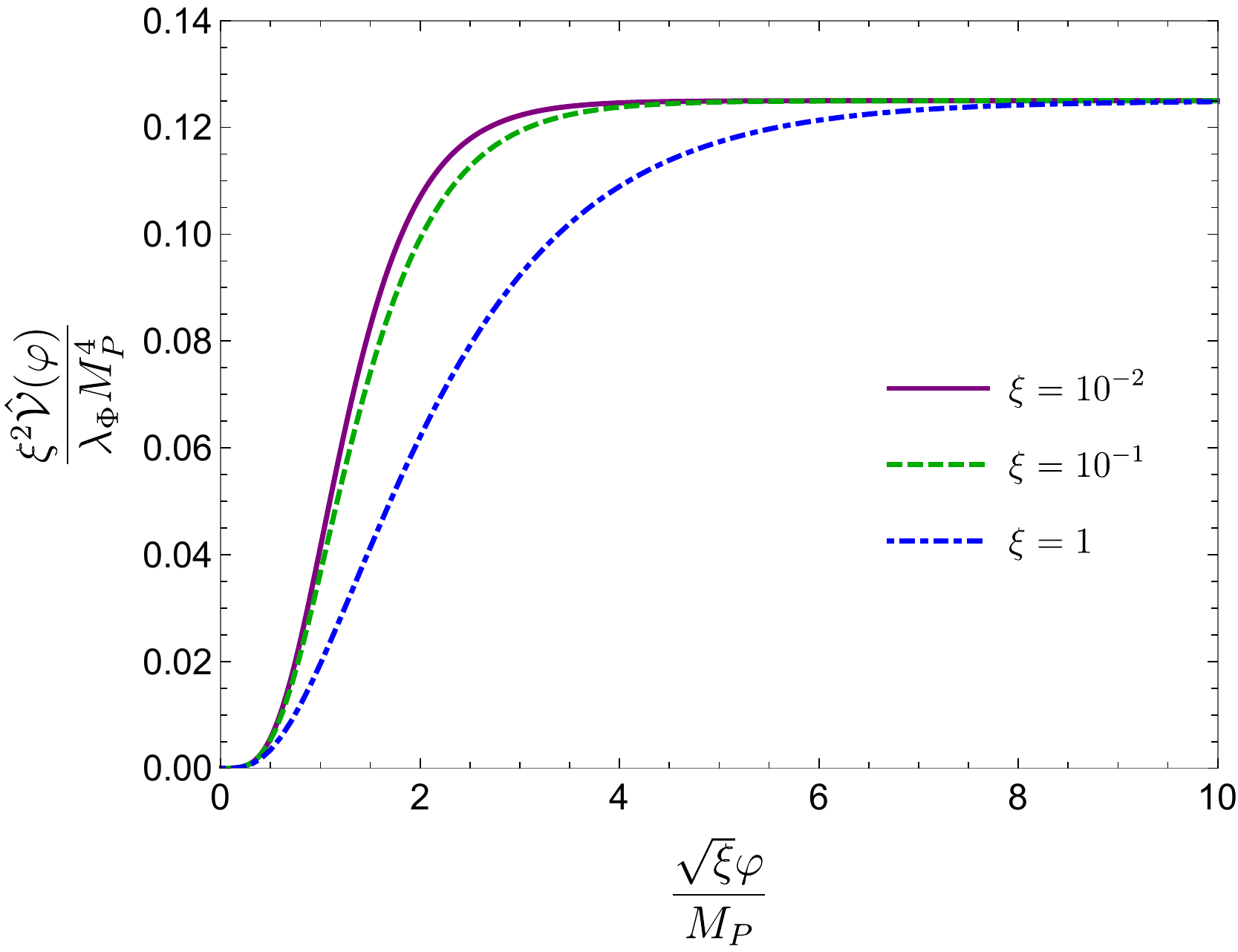}
\caption{
Inflation potential with the non-minimal coupling $\xi=10^{-2}$
$10^{-1}$, and $1$.
\bigskip}
\label{fig:inflatonpotential}
\end{figure}

The non-minimal coupling is removed by the conformal transformation, 
\begin{align}
 g_{\mu\nu} \to g_{\mu\nu} = \Omega^{-2} \hat{g}_{\mu\nu}
 \quad
 \text{with}
 \quad
 \Omega = \sqrt{1 + \frac{\xi \phi^2}{M_P^2}},
\end{align}
where $\hat{g}_{\mu\nu}$ corresponds to the metric in the Einstein frame. 
As a result of this transformation, the Lagrangian becomes
\begin{equation}
 \frac{\hat{\mc{L}}}{\sqrt{-\hat{g}}}=-\frac{M_P^2}{2}\hat{\mc{R}}+\frac{\hat{g}^{\mu\nu}}{2}\left(\partial_{\mu}\varphi\right)\left(\partial_{\nu}\varphi\right)
+\frac{\hat{g}^{\mu\nu}}{2}\left(\partial_{\mu}\hat{\chi}\right)\left(\partial_{\nu}\hat{\chi}\right)
-\hat{\mc{V}}(\varphi),
\end{equation}
where $\varphi$ and $\hat{\chi}$ are the canonically normalized fields
satisfying the differential equations 
\begin{equation}
\frac{d\varphi}{d\phi}=\sqrt{\frac{M_P^2\Omega^2+6\xi^2\phi^2}{M_P^2\Omega^4}},\qquad
\frac{d\hat{\chi}}{d\chi}=\Omega^{-1}.
\label{eq:diff}
\end{equation}
If $\xi\gg1$, which is similar to the case of the Higgs
inflation~\cite{Bezrukov:2007ep}, 
the differential equation is simplified and the explicit expressions of
$\varphi$ and $\hat{\mc{V}}(\varphi)$ are obtained.
However in the present pNGB model with feeble couplings, we will show
later that a smaller $\xi$ is favored for successful inflation.
The shape of the scalar potential is numerically evaluated and shown
in Fig.~\ref{fig:inflatonpotential} where the non-minimal coupling is
fixed to be $\xi=10^{-2},10^{-1},1$.
The flat part of the potential gets longer for a smaller value of $\xi$.

\begin{figure}[t]
\begin{center}
\includegraphics[scale=0.62]{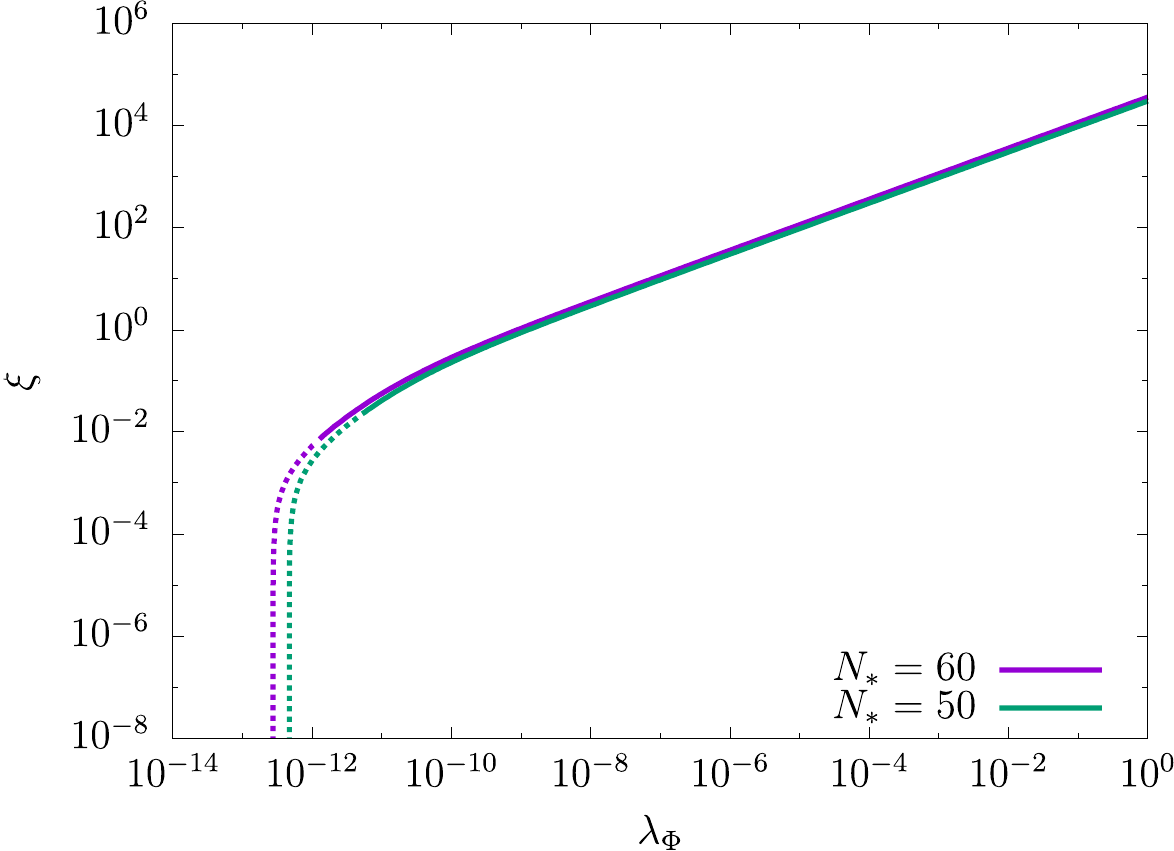} ~
\includegraphics[scale=0.62]{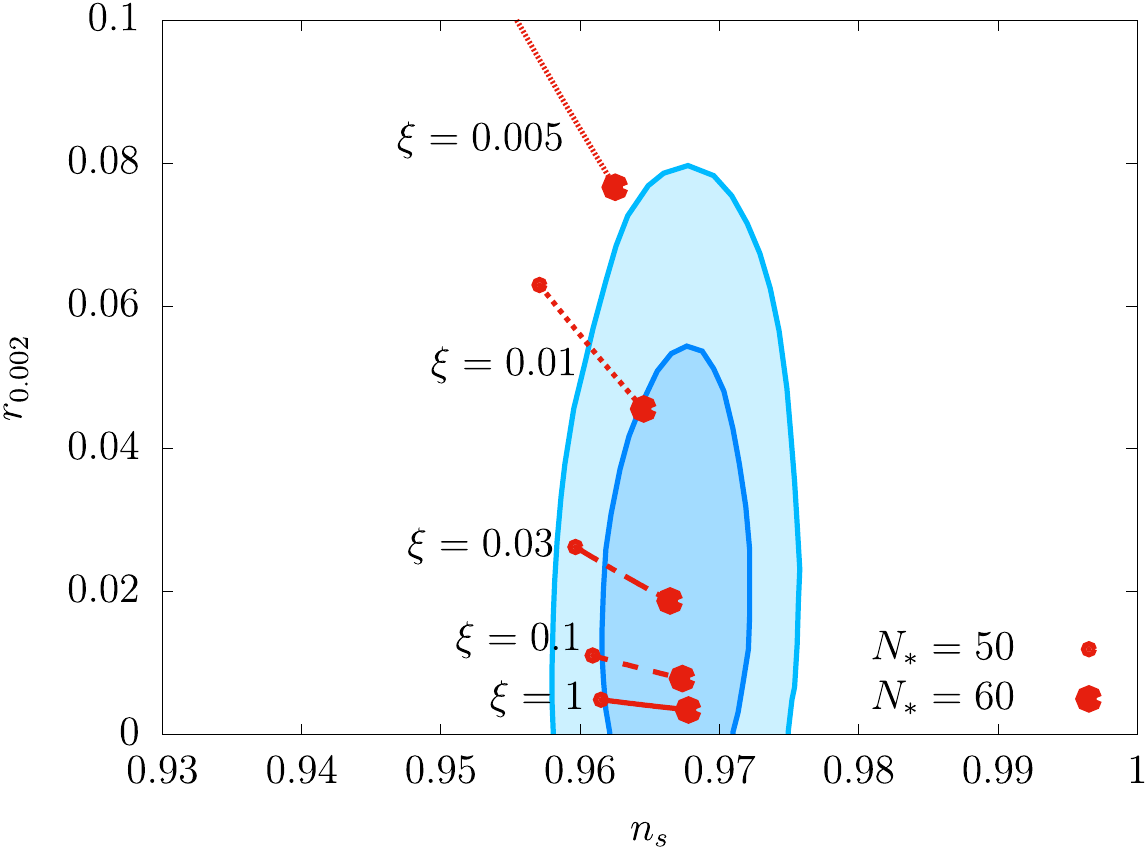}
\caption{(Left): Contours in the ($\xi$, $\lambda_\Phi$) plane for
successful inflation where $N_*$ is fixed as $N_*=50$ and $60$.
(Right): Predictions for the spectral index ($n_s$) and the tensor-to-scalar ratio ($r$) where the non-minimal coupling $\xi$ is taken as $\xi=1$, $0.1$, $0.03$, $0.01$ and $0.005$. The blue and light blue regions are allowed by the PLANCK
observation at $1\sigma$ and $2\sigma$ confidence level, respectively.
\bigskip}
\label{fig:inf}
\end{center}
\end{figure}

While one cannot write down the explicit form of the scalar potential
$\hat{\mc{V}}(\varphi)$, the slow-roll parameters are expressed in terms
of $\phi$,
\begin{align}
\epsilon\equiv &~ \frac{M_P^2}{2}\left(\frac{\hat{\mc{V}}_\varphi}{\hat{\mc{V}}}\right)^2
= \frac{8M_P^4}{\phi^2[M_P^2+\xi\left(1+6\xi\right)\phi^2]},\label{eq:eps}\\
\eta\equiv &~ M_P^2\frac{\hat{\mc{V}}_{\varphi\varphi}}{\hat{\mc{V}}}
=\frac{4M_P^2[3M_P^4+M_P^2\xi\left(1+12\xi\right)\phi^2-2\xi^2\left(1+6\xi\right)\phi^4]}{\phi^2\left[M_P^2+\xi\left(1+6\xi\right)\phi^2\right]^2},
\end{align}
where $\hat{\mc{V}}_\varphi\equiv\partial\hat{\mc{V}}/\partial\varphi$ and $\hat{\mc{V}}_{\varphi\varphi}\equiv\partial^2\hat{\mc{V}}/\partial\varphi^2$.
Using the slow-roll approximation, the spectral index $n_s$ and the
tensor-to-scalar ratio $r$ are given by 
\begin{align}
 n_s=1-6\epsilon+2\eta, \qquad
 r=16\epsilon.
\end{align}
The e-folding number $N_*$ between the time of horizon exit ($t_*$)
and the end of inflation ($t_\mathrm{end}$) is given by 
\begin{align}
N_*=\int_{t_*}^{t_\mr{end}}\!Hdt
\,\approx\,\frac{-1}{M_P^2}\int^{\phi_\mr{end}}_{\phi_*}\frac{d\varphi}{d\phi}\frac{\hat{\mc{V}}}{\mc{\hat{V}}_\varphi}d\phi
\,=\frac{\left(1+6\xi\right)\left(\phi_*^2-\phi_\mr{end}^2\right)}{8M_P^2}-\frac{3}{4}\log\left(\frac{M_P^2+\xi\phi_*^2}{M_P^2+\xi\phi_\mr{end}^2}\right),
\end{align}
where $\phi_*$ and $\phi_\mathrm{end}$ are the field values at $t_*$
and $t_\mathrm{end}$, respectively. 
Defining $t_\mathrm{end}$ as the time giving the slow-roll parameter
$\epsilon=1$ in Eq.~(\ref{eq:eps}), $\phi_\mathrm{end}$ satisfies 
\begin{equation}
 \phi_\mr{end}^2=\frac{M_P^2}{2}\,\frac{\sqrt{1+32\xi+192\xi^2}-1}{\xi\left(1+6\xi\right)}.
\end{equation}
On the other hand, $\phi_*$ is numerically evaluated,  
giving the amplitude of the scalar power spectrum at the horizon exit 
observed by the Planck Collaboration~\cite{Akrami:2018odb}: 
$A_s=\hat{\mc{V}}(\varphi(\phi_*))/\left(24\pi^2 M_P^4\,\epsilon(\phi_*)\right)=2.10\times10^{-9}$. From these
relations, the non-minimal coupling $\xi$ and the scalar self-coupling $\lambda_\Phi$ for successful inflation can be read.
The contour of the e-folding number $N_*$ on the ($\xi$, $\lambda_\Phi$) plane is shown in the left panel of Fig.~\ref{fig:inf}
where $N_*=50$ and $60$. 
When the non-minimal coupling $\xi$ is small enough
($\xi\lesssim10^{-3}$), the dependence on $\lambda_\Phi$ disappears. This
region has the same behavior as the $\phi^4$ chaotic inflation.

The predictions for the spectral index $n_s$ and the tensor-to-scalar ratio $r$ at the pivot scale $k_*=0.002~\mathrm{Mpc}^{-1}$ 
are shown in the right panel of Fig.~\ref{fig:inf} where $N_*$ is
taken to be between $50$ and $60$. 
The blue and light blue regions represent the $1\sigma$ and $2\sigma$ confidence levels observed by the Planck Collaboration~\cite{Akrami:2018odb}.
It can be found that the lower bounds for the non-minimal coupling are required
\begin{align}
 \xi \gtrsim 0.02 \quad (N_*=50), \qquad 
 \xi \gtrsim 0.0055  \quad (N_*=60),
\end{align}
for the scenario consistent with the observation.

In the left panel of Fig.~\ref{fig:inf}, the solid (dotted) part of
each contour represents the parameter region consistent (inconsistent)
with the PLANCK observation at $2\sigma$ confidence level, when
combined with the lower bounds of $\xi$ obtained in the right panel. 
We find that the self-coupling $\lambda_\Phi$ should be in the range 
$\lambda_\Phi \gtrsim 10^{-12}$
in the pNGB dark matter model with large symmetry breaking,
if the inflation is induced by the coupling $\xi$.

\subsection{PNGB production from inflaton}

There is another important physical implication of the possibility that
the radial scalar component $\phi$ plays the role of inflaton 
in the pNGB dark matter model.
That is the direct production process of dark matter $\chi$ from the inflaton
decay, which is inevitable because of the interaction between $\phi$ and $\chi$.
In this section, we investigate the parameter space consistent with the inflationary scenario discussed above and 
the dark matter relic abundance taking into account both the freeze-in and
inflaton-induced dark matter.

As discussed in Refs.~\cite{Takahashi:2007tz, Gorbunov:2010bn},
the number density of the inflaton induced $\chi$ particle is estimated as
\begin{align}
 n_\chi^{\mr{inf}} \approx \frac{\rho_{\rm rh}}{m_\phi} \frac{\Gamma_{\phi \to \chi \chi}}{\sqrt{3} H(T_R)},
\end{align}
where $\rho_{\rm rh}$ and $H(T_R)$ are the energy density of radiation
and the Hubble rate at the reheating temperature $T_R$. 
We here define the reheating temperature $T_R$ 
at which the Hubble rate is equal to the decay width of the inflaton
to the SM sector ($H(T_R)\approx\Gamma_{\phi\to H^{\dag}H}$), and then
\begin{align}
 T_R^2\, \approx \frac{3\sqrt{10}}{8\pi^2 g_*^{1/2}} 
 \frac{\lambda_{H\Phi}^2}{\lambda_\Phi}M_P m_\phi
  \sqrt{1 - \frac{4 m_h^2}{m_\phi^2}}.
 \label{eq:tr}
\end{align}
Note that the effective degrees of freedom $g_*$ in the right-hand
side also depends on the temperature in general. 
Using the explicit form of the decay widths and the reheating
temperature \eqref{eq:tr}, we obtain the pNGB dark matter yield
directly produced from the inflaton as
\begin{align}
 Y_\chi^{\mr{inf}}
 \approx \biggl(\frac{9\sqrt{10}}{2048\pi^2}\biggr)^{1/2}
\frac{g_*^{3/4}\lambda_\Phi^{3/2}M_P^{1/2}}{g_*^S\lambda_{H\Phi}\,m_\phi^{1/2}}
 \biggl(1 - \frac{4 m_\chi^2}{m_\phi^2}\biggr)^{\frac{1}{2}}
 \biggl(1 - \frac{4 m_h^2}{m_\phi^2}\biggr)^{-\frac{1}{4}}.
 \label{eq:Yinf}
\end{align}
As previously, we impose the portal coupling 
satisfies $\lambda_{H\Phi} \lesssim 10^{-6}$
such that dark matter does not get into the SM thermal bath.

\begin{figure}[t]
\centering
\includegraphics[scale=0.63]{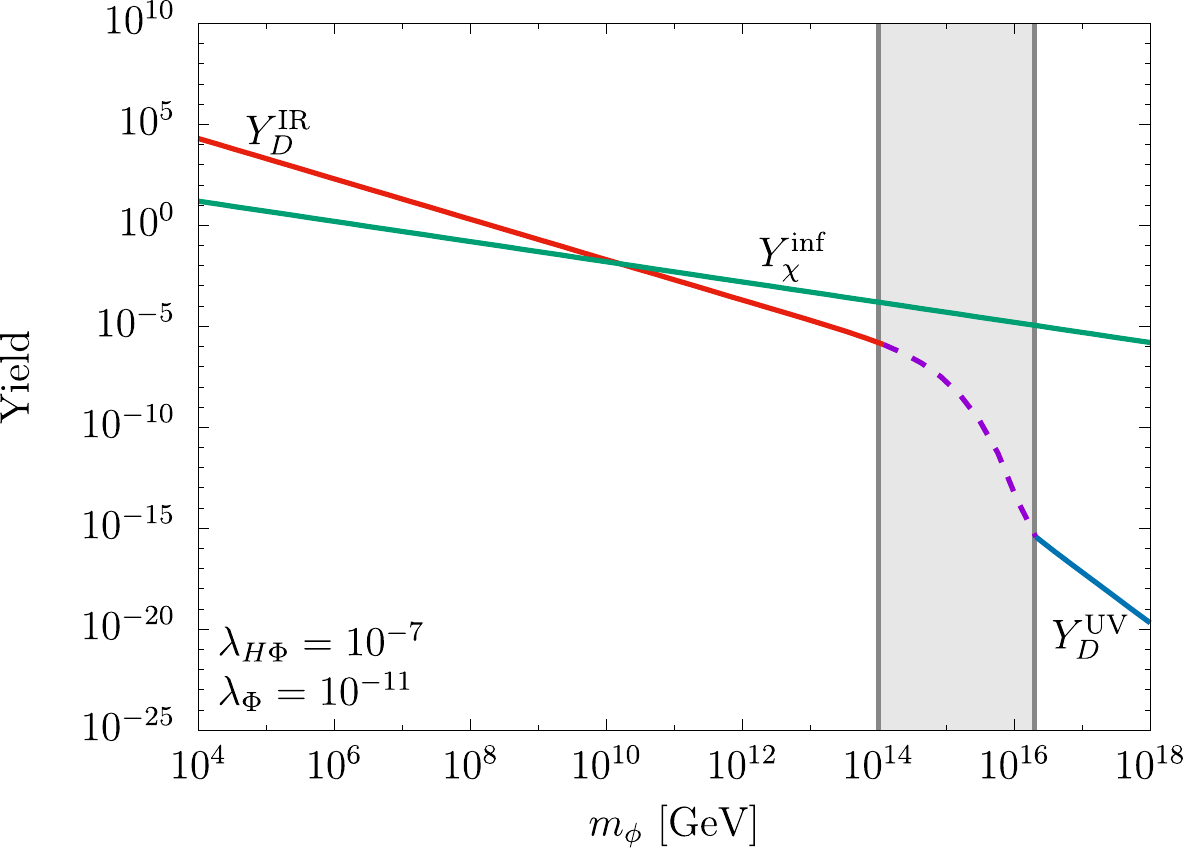} ~
\includegraphics[scale=0.63]{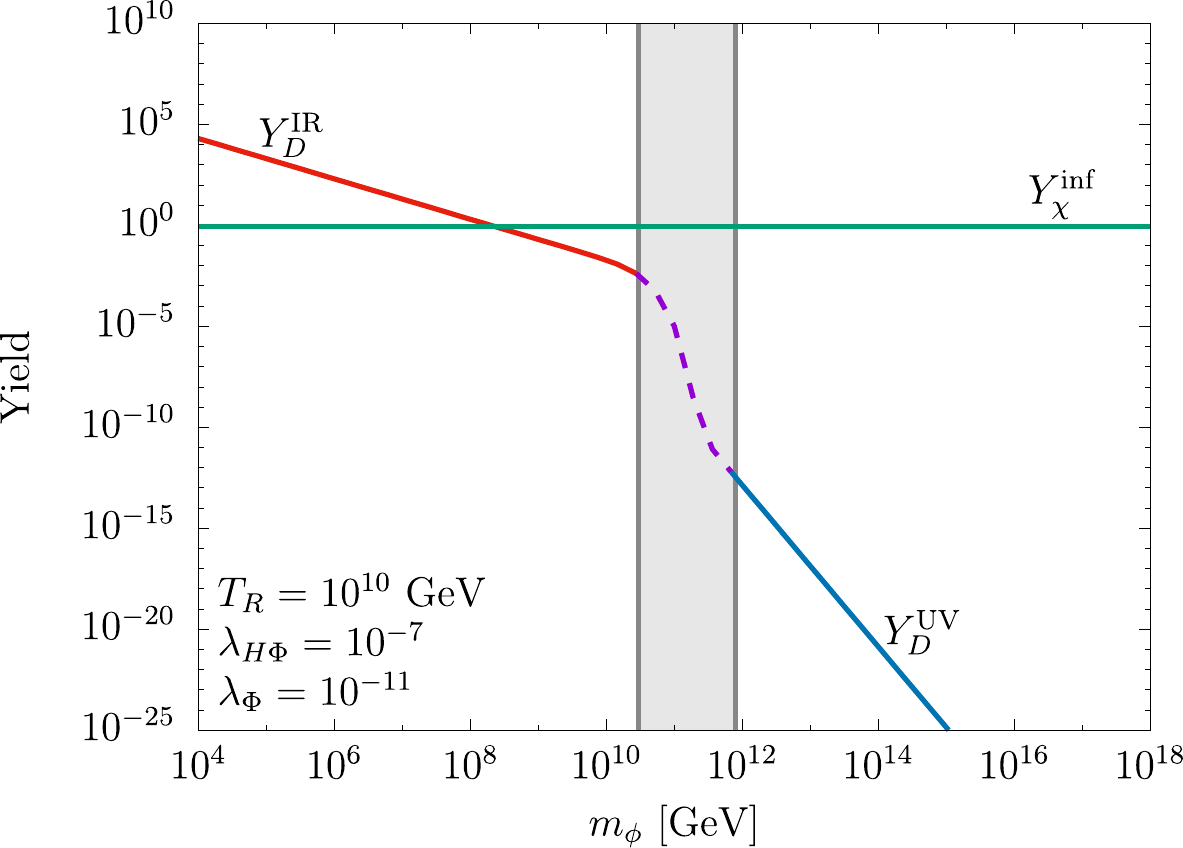}
\caption{
Comparisons of the pNGB dark matter yields, $Y^{\mr{IR}}_D$ 
(red), $Y^{\mr{UV}}_D$ (green) and $Y^{\mr{inf}}_\chi$ (blue) 
for $\lambda_{H\Phi} = 10^{-7}$ and $\lambda_\Phi=10^{-11}$.
The reheating temperature is given by Eq.~\eqref{eq:tr} in the left panel
and a free parameter in the right panel, 
chosen as $T_R = 10^{10}~\mr{GeV}$.
The IR and UV freeze-in are smoothly connected in the gray region,
typically $T_R \lesssim m_\phi \lesssim 100\, T_R$.
\bigskip}
\label{fig:YIRUVinf}
\end{figure}

We here comment on the comparison between the three contributions of
pNGB dark matter yields, $Y^{\mr{IR}}_D$, $Y^{\mr{UV}}_D$ and $Y^{\mr{inf}}_\chi$.
Typical behaviors are shown in Fig.~\ref{fig:YIRUVinf} as
the functions of the mediator mass $m_\phi$, which behaviors 
are evaluated by MicrOMEGAs 
for $\lambda_{H\Phi}= 10^{-7}$ and $10^{-11}$. 
The reheating temperature is given by Eq.~\eqref{eq:tr} in the left
panel, while it is treated as a free parameter in the right panel (see
the detail in the next subsection).
The IR and UV freeze-in productions are effective in the smaller
and larger $m_\phi$ regions, respectively.
These freeze-in yields are smoothly connected in the gray region 
where $T_R \lesssim m_\phi \lesssim 100 \, T_R$,
which is denoted by the purple dashed line.
Since $Y^\mr{IR}_D\propto m_\phi^{-1}$ and 
$Y^\mr{inf}_\chi\propto m_\phi^{-1/2}$, 
the IR freeze-in is dominant for a lighter mediator. Further,
$Y^\mr{IR}_D$ becomes equal to $Y^\mr{inf}_\chi$ at $m_\phi=m_\mr{eq}$. It is
easy to find from 
Eqs.~\eqref{eq:relicapprox}, \eqref{eq:tr} and \eqref{eq:Yinf} that 
$m_\mr{eq}\lesssim 10^{-2}\,T_R$, and
then the inflaton-induced yield necessarily becomes 
dominant in the left side of the gray band.
On the other hand, the UV freeze-in abundance takes the value
$Y^\mr{UV}_D/Y^\mr{inf}_\chi \simeq 10^{-3}\lambda_{H\Phi}^2/\lambda_\Phi$ around
 $m_\phi=T_R$. This ratio is much smaller than 1 due to the 
constraints $\lambda_{H\Phi}\lesssim10^{-6}$ and 
$\lambda_\Phi\gtrsim10^{-12}$ as we explained from the cosmological arguments.
Since $Y^\mr{UV}_D\propto m_\phi^{-5/2}$,
the UV freeze-in contribution is always subdominant compared to 
the inflaton decay.

\subsection{Dark matter abundance}

\begin{figure}
\centering
\includegraphics[scale=0.427]{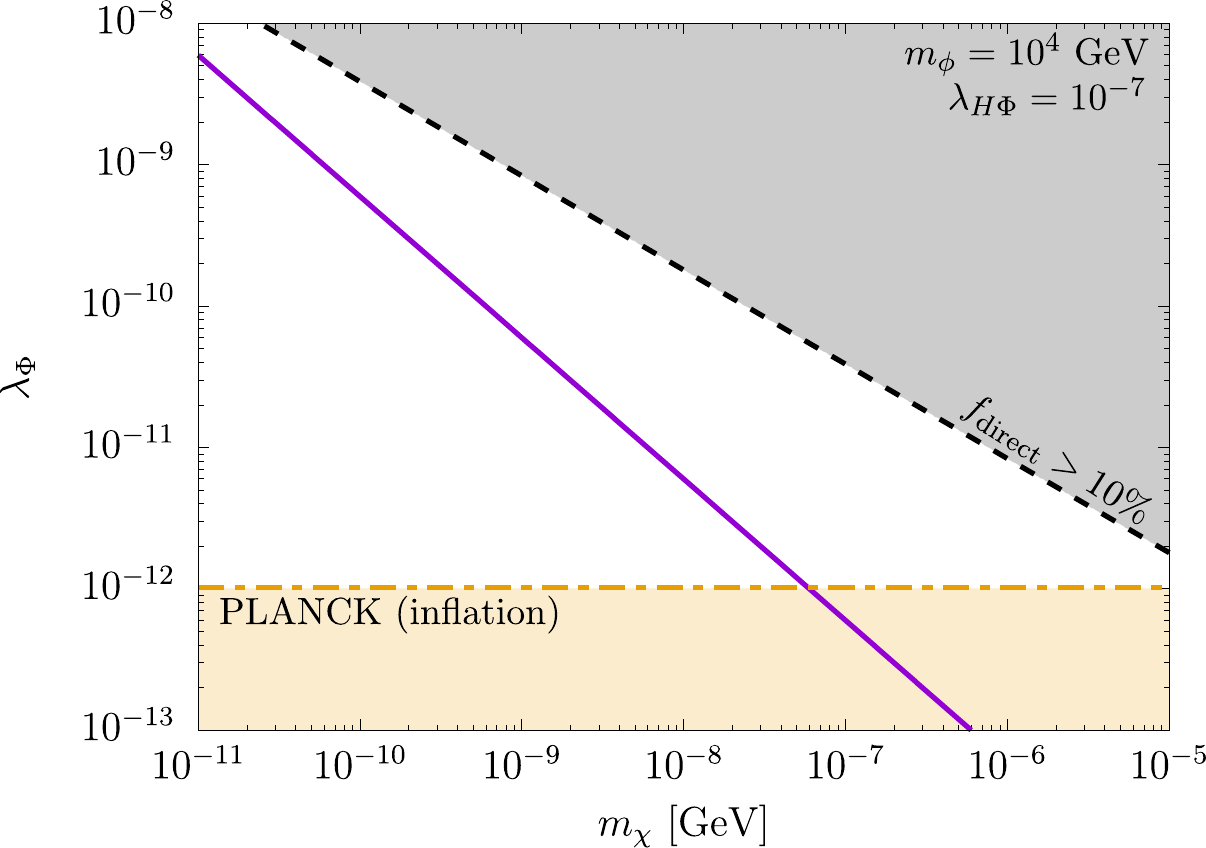}
\includegraphics[scale=0.427]{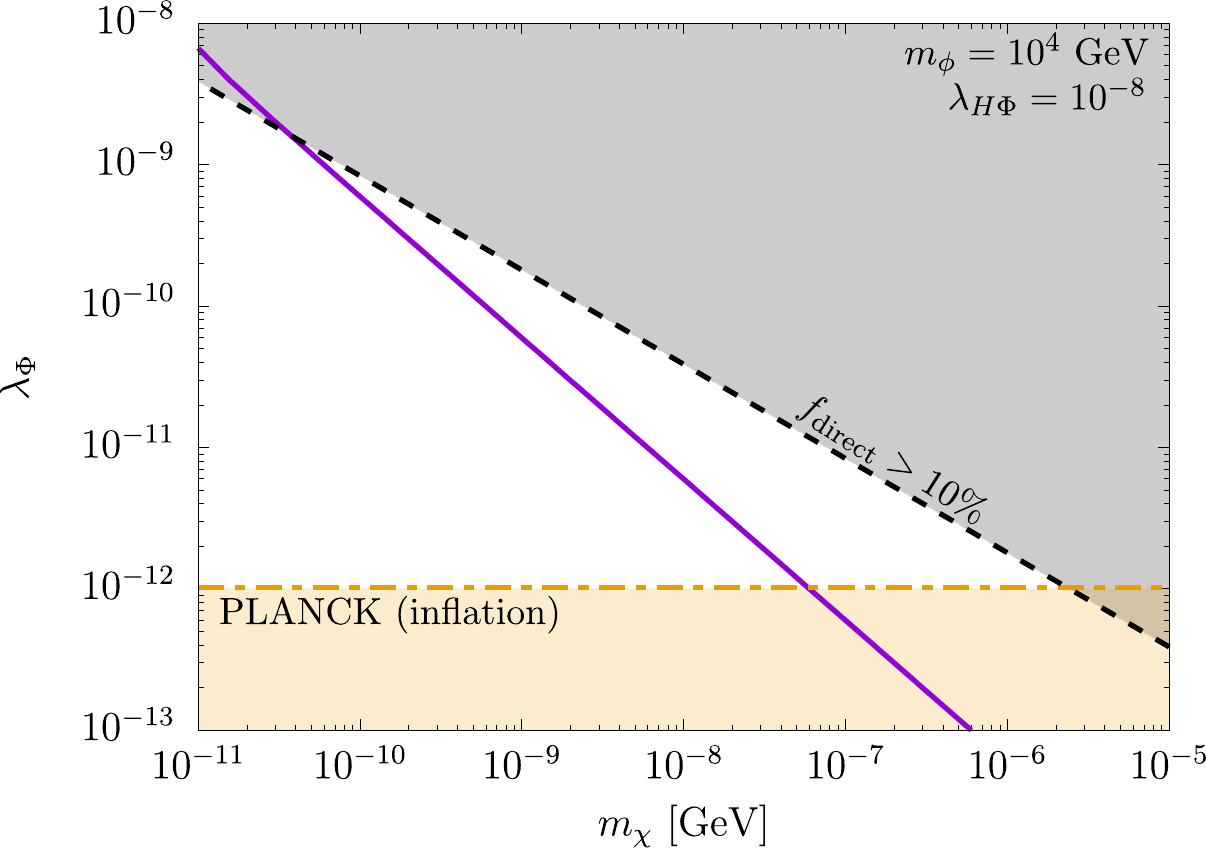}
\includegraphics[scale=0.427]{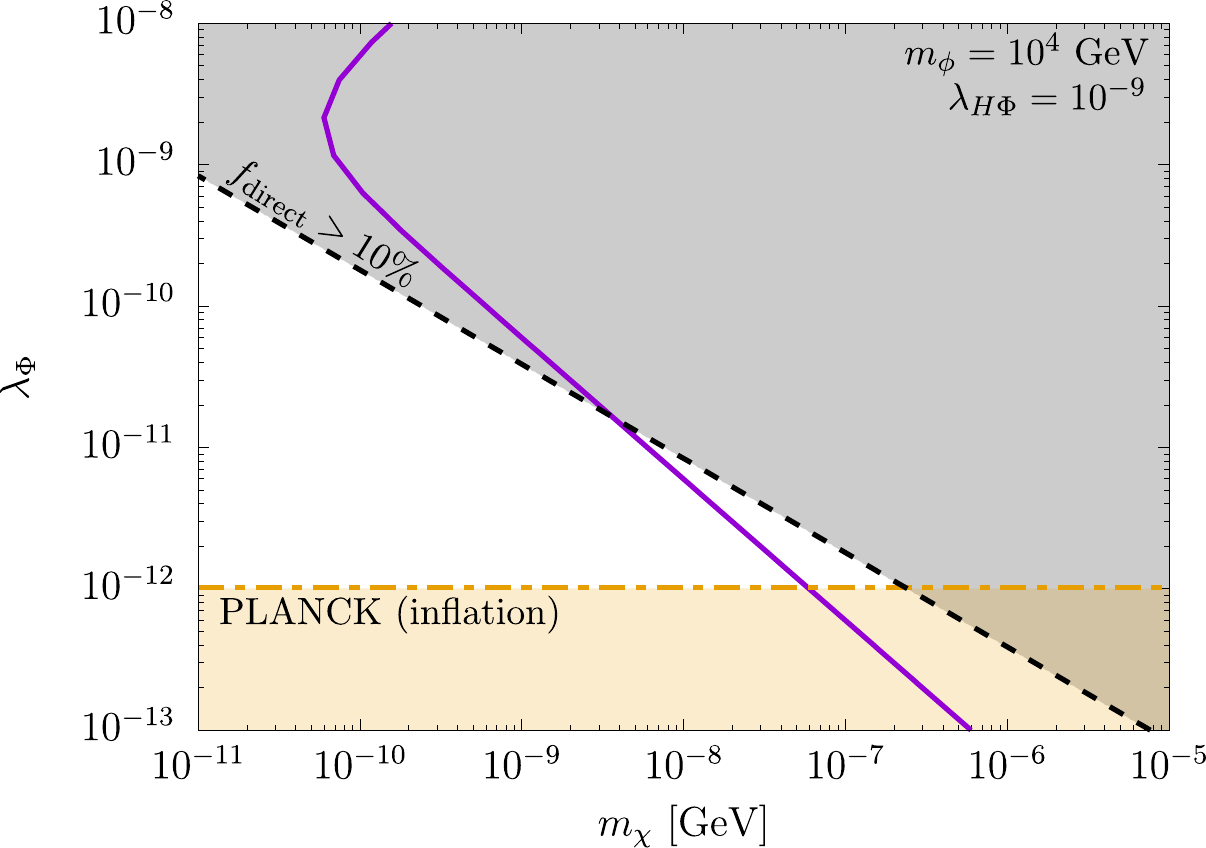}
\\
\includegraphics[scale=0.427]{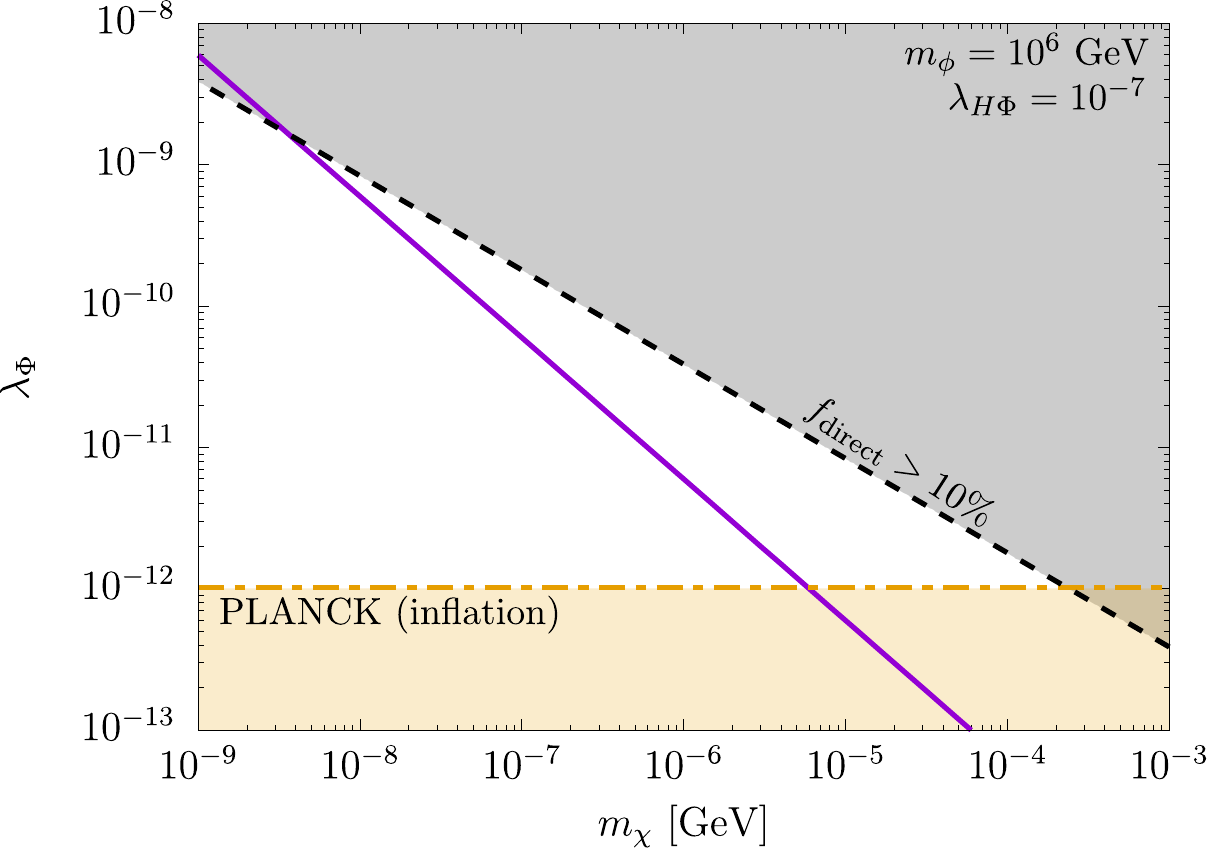}
\includegraphics[scale=0.427]{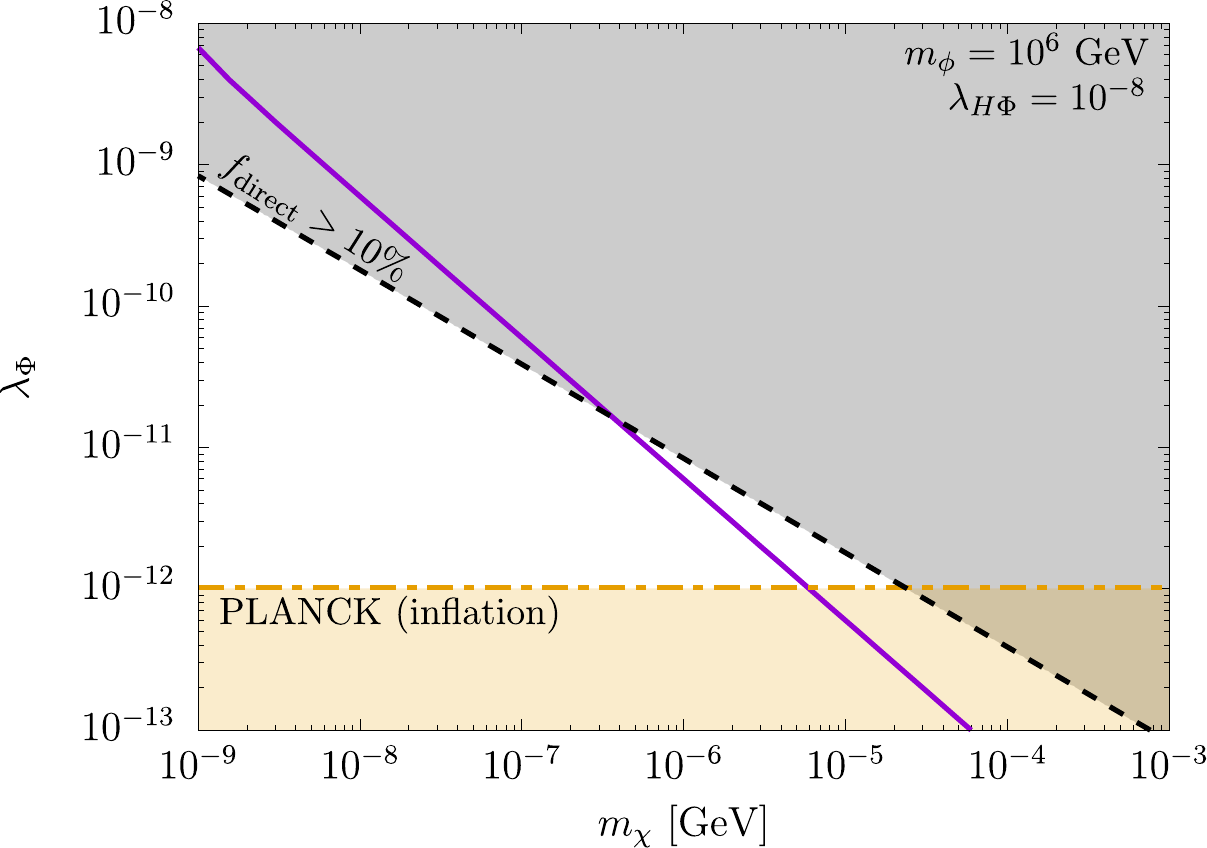}
\includegraphics[scale=0.427]{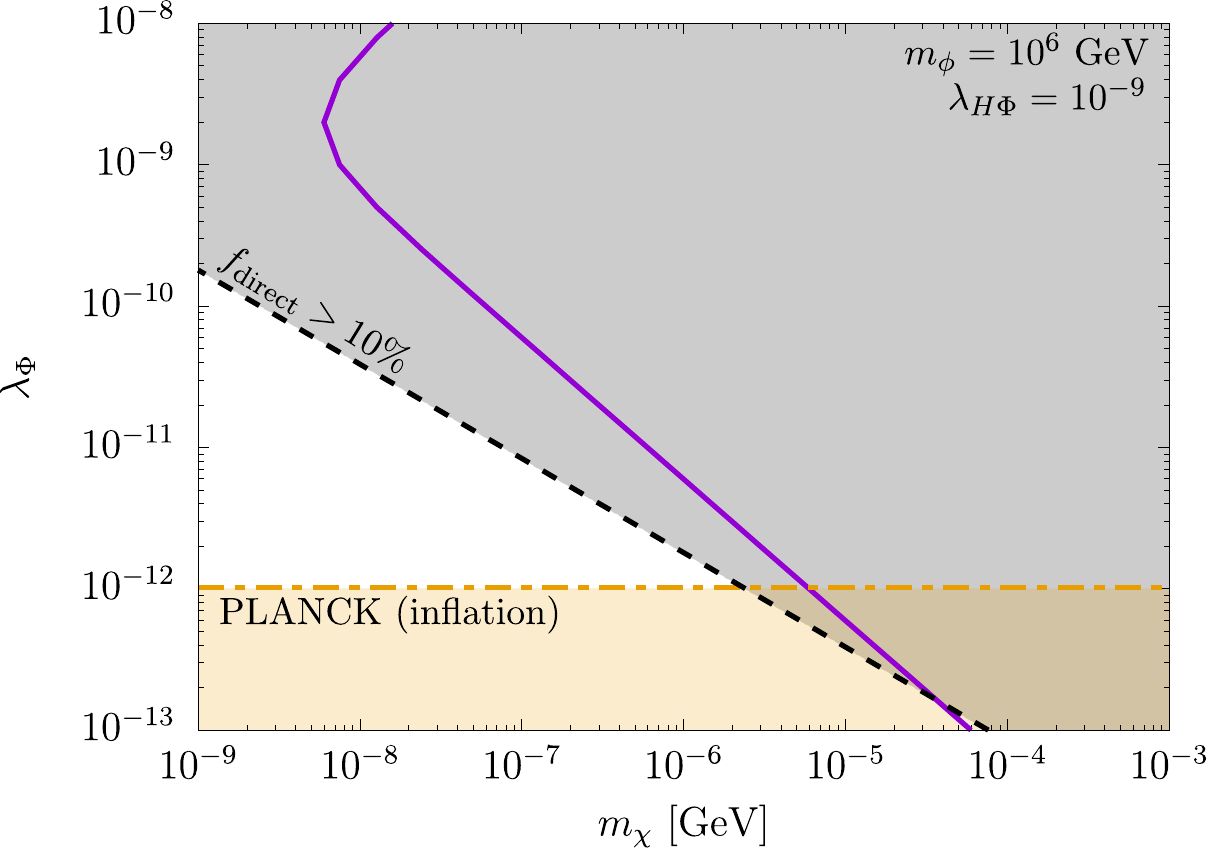}
\\
\includegraphics[scale=0.427]{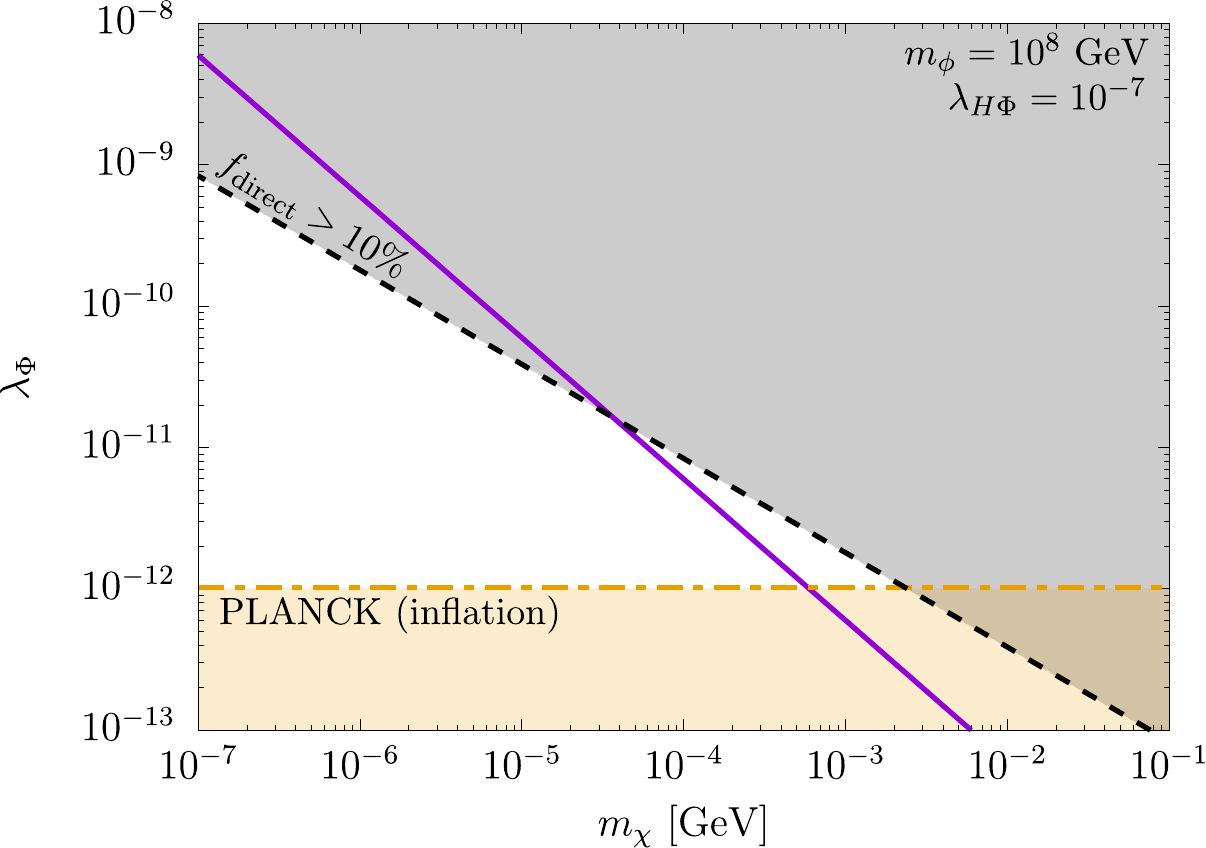}
\includegraphics[scale=0.427]{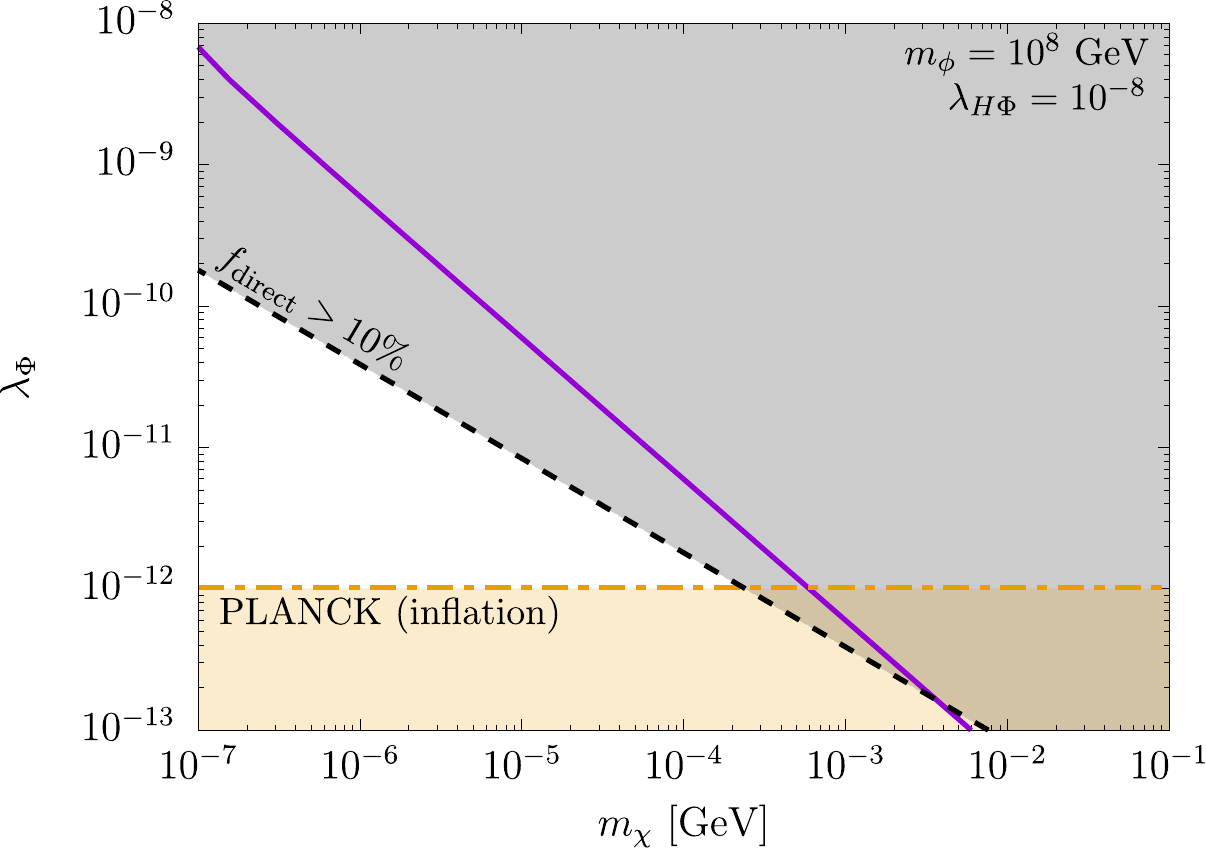}
\includegraphics[scale=0.427]{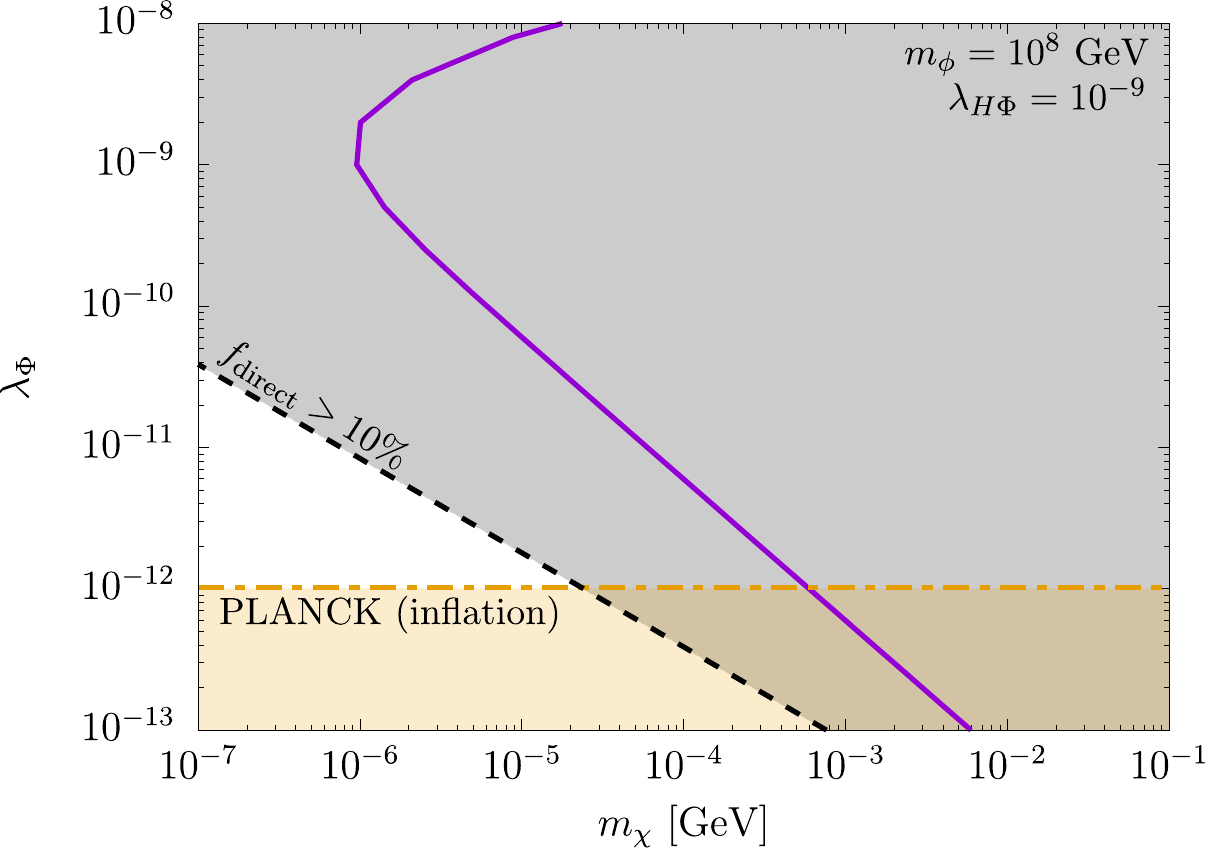}
\caption{
Parameter space in the $(m_\chi, \lambda_\Phi)$ plane
where the reheating of the universe is assumed to occur via the
perturbative inflaton decay into the SM sector.
The purple lines reproduce the correct relic abundance of dark matter
only by the freeze-in production.
The gray regions represent the parameter space that the dark matter
abundance created by the inflaton decay is larger than $10~\%$ of the observed value.
The lower orange region cannot be consistent with the inflation observables
at $2 \sigma$ confidence level.
\bigskip}
\label{fig:allowedm_chi-lambda_Phi}
\end{figure}

Fig.~\ref{fig:allowedm_chi-lambda_Phi} shows the parameter space 
realizing the correct dark matter relic abundance in 
the $(m_\chi,\lambda_\Phi)$ plane 
taking into account the pNGB production from the freeze-in and inflaton decay.
The portal coupling is chosen as $\lambda_{H\Phi} = 10^{-7}$, $10^{-8}$ and $10^{-9}$ and the mediator mass is
$m_\phi = 10^{4}~\mr{GeV}$, $10^{6}~\mr{GeV}$ and $10^{8}~\mr{GeV}$.
The purple line represents the parameters reproducing the dark matter
abundance only by the freeze-in contribution.
The gray region represents the parameter space where the inflaton
induced dark matter abundance is larger than $10~\%$ of the observed value.
In order to be consistent with the inflation observables at $2 \sigma$
confidence level as discussed in the previous subsection, 
the scalar self-coupling has to satisfy $\lambda_\Phi \gtrsim 10^{-12}$ and 
the lower orange region is excluded by this condition.

The behavior of the purple lines in the most of panels can be understood by the IR freeze-in as discussed in Sec.~\ref{sec:IRFI}, except the right-bottom one.
The region $\lambda_{\Phi}\gtrsim10^{-9}$ of that plot corresponds to
the intermediate state between the IR and UV freeze-in, namely
$T_R\sim m_{\phi}$, while the contribution to the relic from the
inflaton decay is eventually dominant in this region. 
In case that the reheating temperature determined by the inflaton
decay is much lower than the mediator mass $m_{\phi}$, since the
reheating temperature scales as
$T_R\propto\lambda_{H\Phi}\lambda_{\Phi}^{-1/2}m_{\phi}^{1/2}$ which
can be seen in Eq.~(\ref{eq:tr}), it finds that the self-coupling goes
as $\lambda_{\Phi}\propto m_{\chi}^{2/3}$ to reproduce the correct
relic abundance from Eq.~(\ref{eq:uv5}) for the UV freeze-in. This
behavior has been numerically checked in our computation.

\begin{figure}[t]
\centering
\includegraphics[scale=0.27]{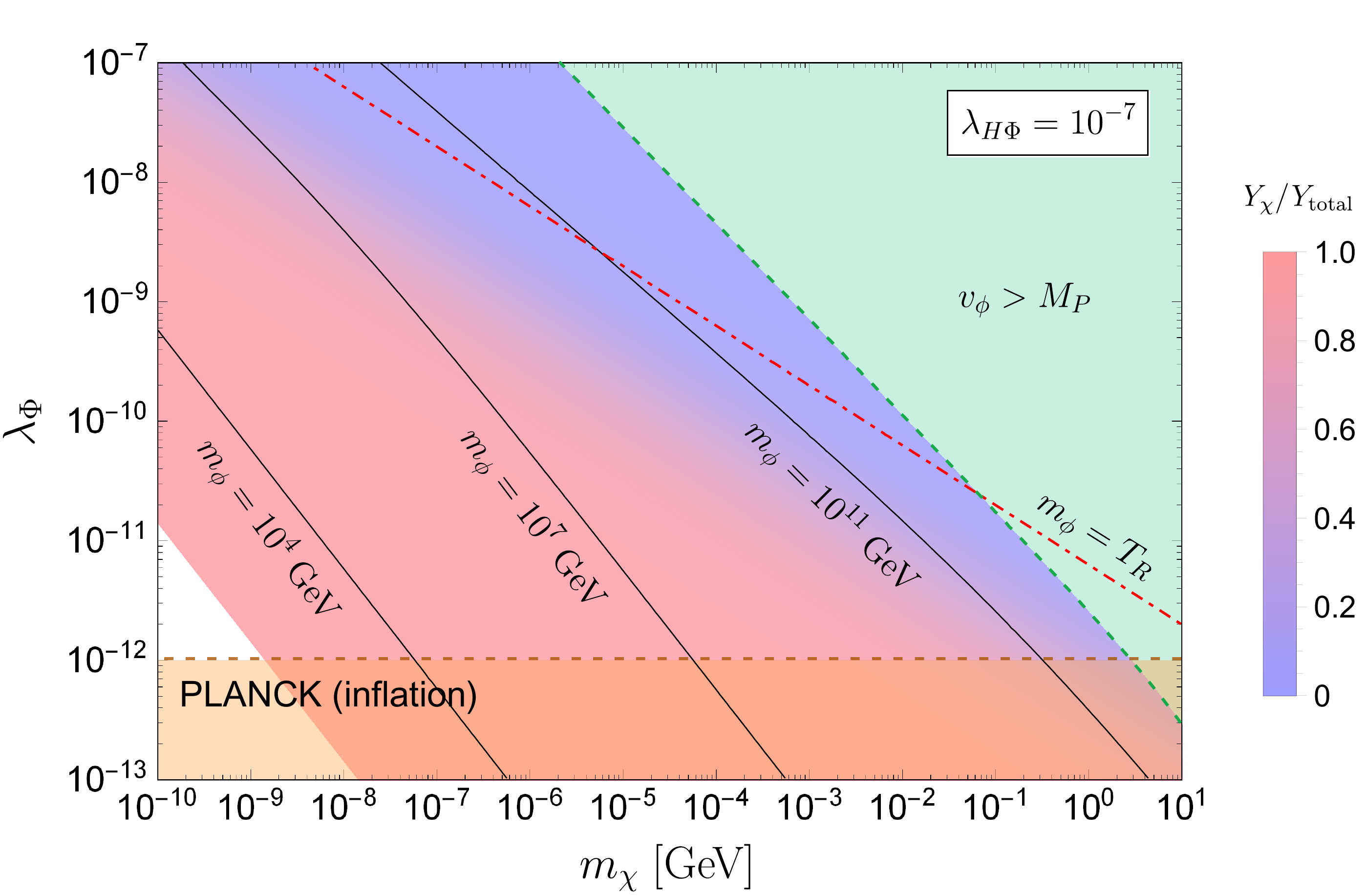}~
\includegraphics[scale=0.27]{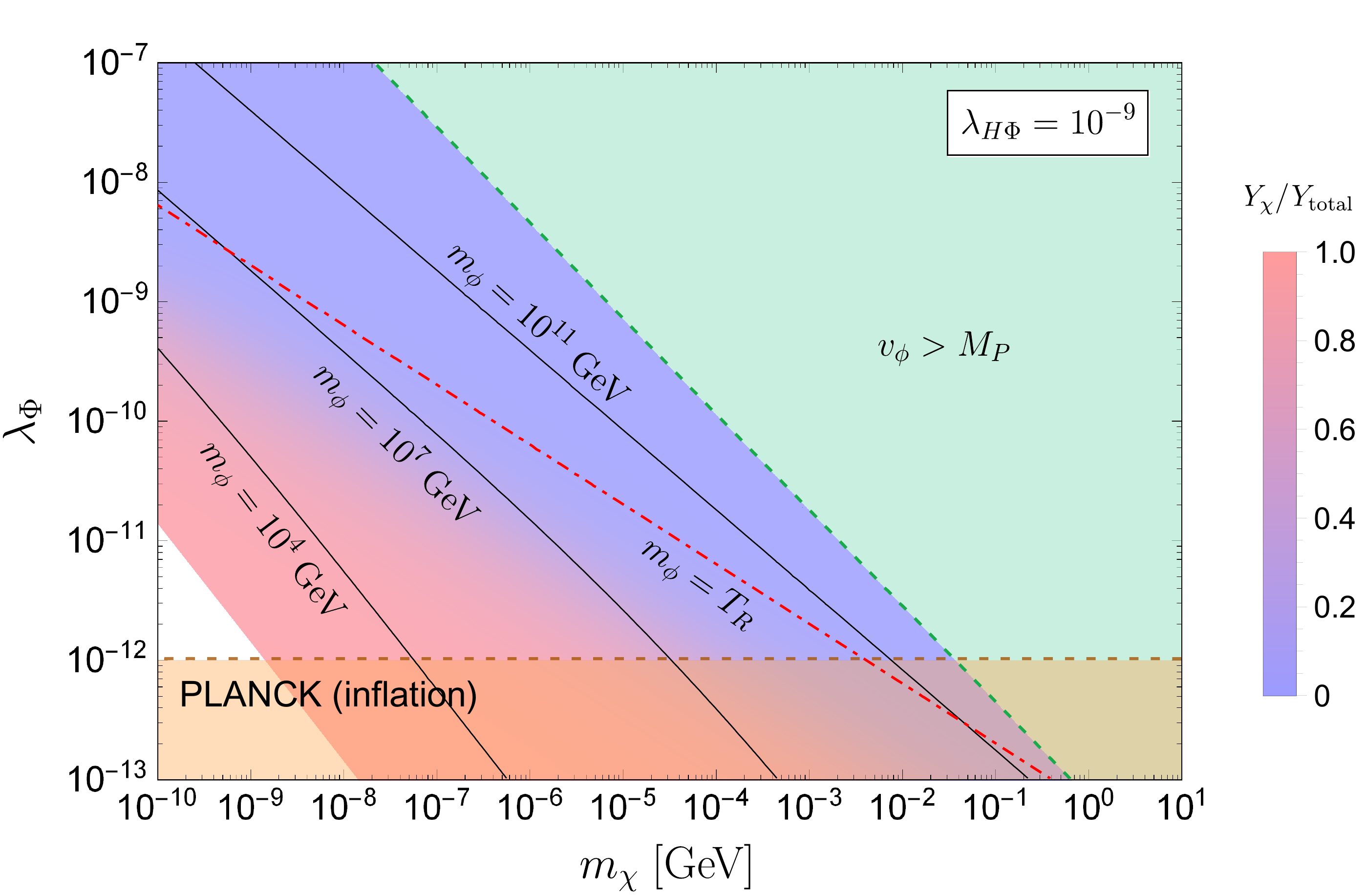}
\caption{
Parameter space in the $(m_\chi, \lambda_\Phi)$ plane
where the correct relic abundance of dark matter is realized by the
non-thermal production (the freeze-in and the inflaton decay). 
The portal coupling is chosen as
$\lambda_{H\Phi} =10^{-7}$ (left) and $\lambda_{H\Phi} = 10^{-9}$ (right).
The freeze-in production is dominant in the red region and the
inflaton decay dominant in the blue region.
The lower orange region is not consistent with the inflation observables as in Fig.~\ref{fig:allowedm_chi-lambda_Phi}.
In the green region, the scalar VEV becomes trans-Planckian.
\bigskip}
\label{fig:allowedm_chi-lambda_Phi2}
\end{figure}

Fig.~\ref{fig:allowedm_chi-lambda_Phi2} shows
the parameter space in the $(m_\chi, \lambda_\Phi)$ plane
realizing the correct dark matter relic abundance
via both contributions from
the freeze-in and inflaton induced productions.
The red and blue regions mean
the abundance is dominated by the freeze-in and the inflaton, respectively. 
The red dot-dashed line denotes $m_\phi = T_R$.
In the green region, the singlet scalar VEV ($v_\phi$) becomes trans-Planckian,
where the low-energy field description is not valid.\footnote{
When one considers graviton loop effect,
its form may be $\frac{m^2}{16\pi^2 M_P^2}$
where $m$ is a typical scale in low-energy theory. Therefore the
low-energy perturbative description is violated above 
$m_P\simeq 4\pi M_P$. If one imposes the condition $v_\phi < m_P$ 
in Fig.~\ref{fig:allowedm_chi-lambda_Phi2},
the green excluded region is relaxed by $4\pi$.
}
\ 
When the dark sector scalars $\phi$ and/or $\chi$ are heavy, the inflaton
decay tends to be the dominant process for the dark matter creation.
A physical implication of Fig.~\ref{fig:allowedm_chi-lambda_Phi2} is that
in the pNGB dark matter model with large symmetry breaking,
the dark matter should be lighter than MeV--GeV 
if the freeze-in production is assumed to be dominant.\footnote{
The light dark matter is constrained from the dark matter free streaming length
being inconsistent with measurements of the Lyman-alpha forest \cite{Viel:2013fqw,Baur:2015jsy}.
This typically excludes the dark matter mass region lighter than keV,
which gives the lower mass bound.
}
\
A heavier dark
matter is also possible if taking the inflaton induced contribution into account.

\begin{figure}
\centering
\includegraphics[scale=0.27]{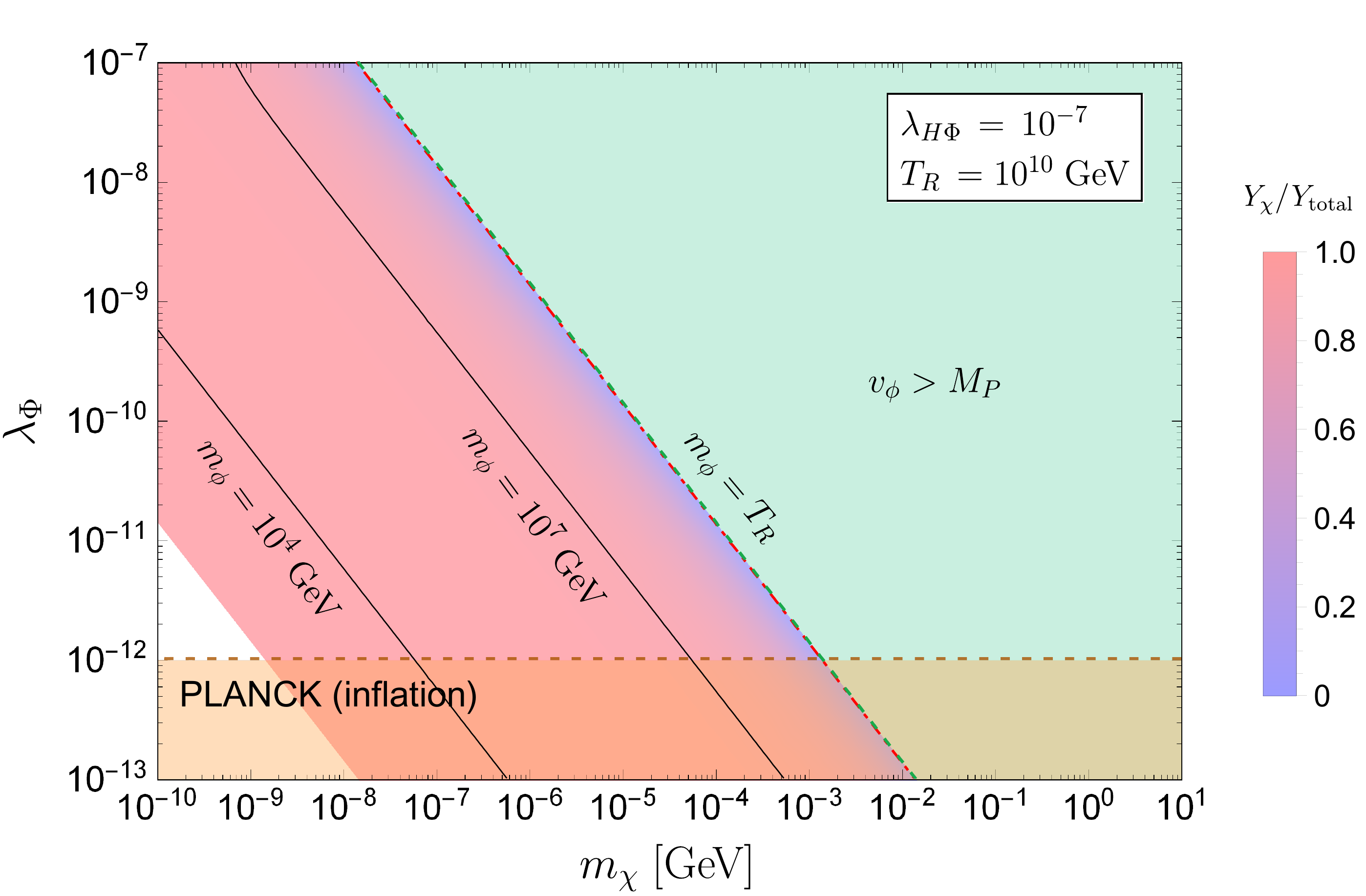}~
\includegraphics[scale=0.27]{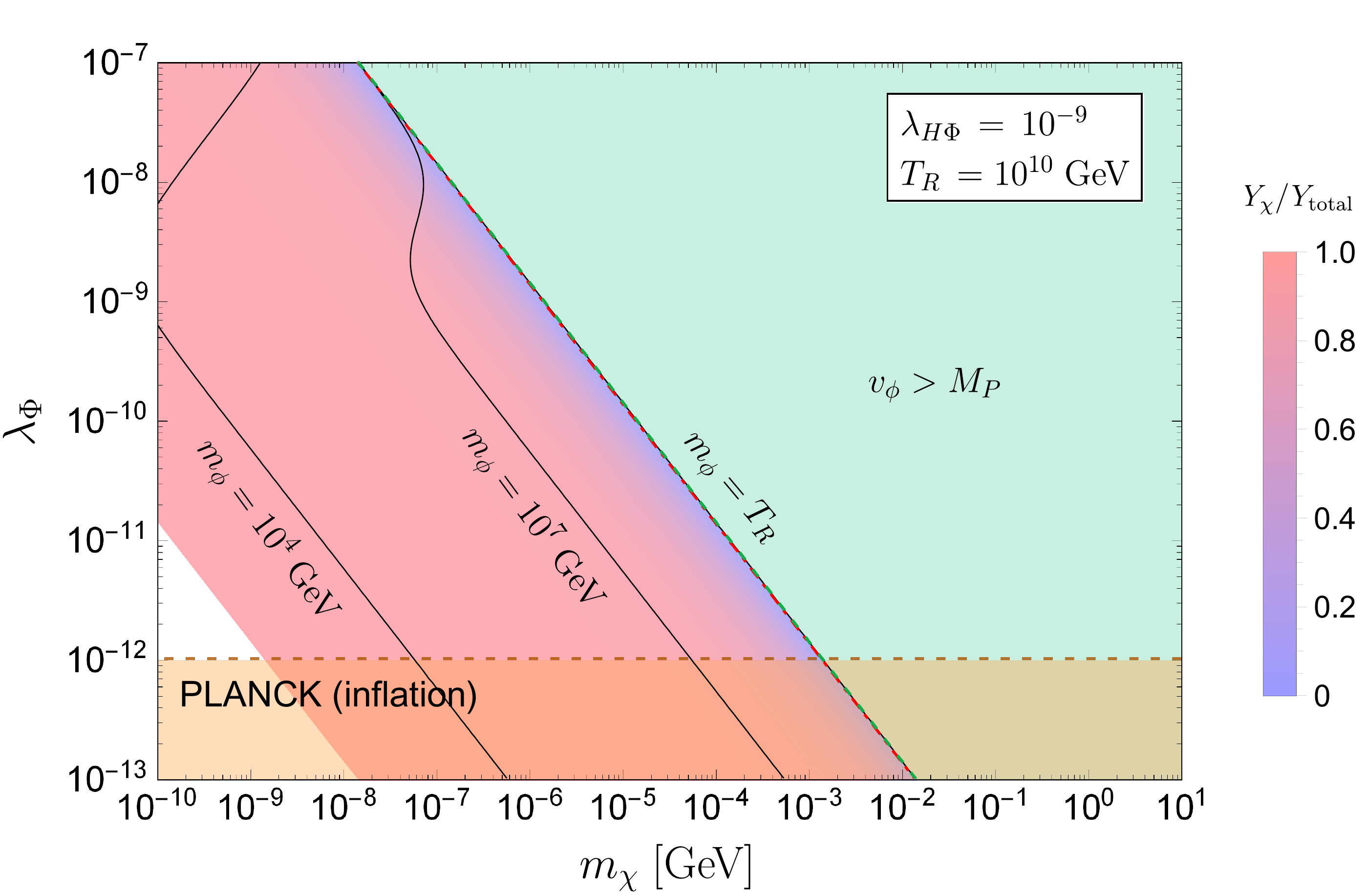}\\
\includegraphics[scale=0.27]{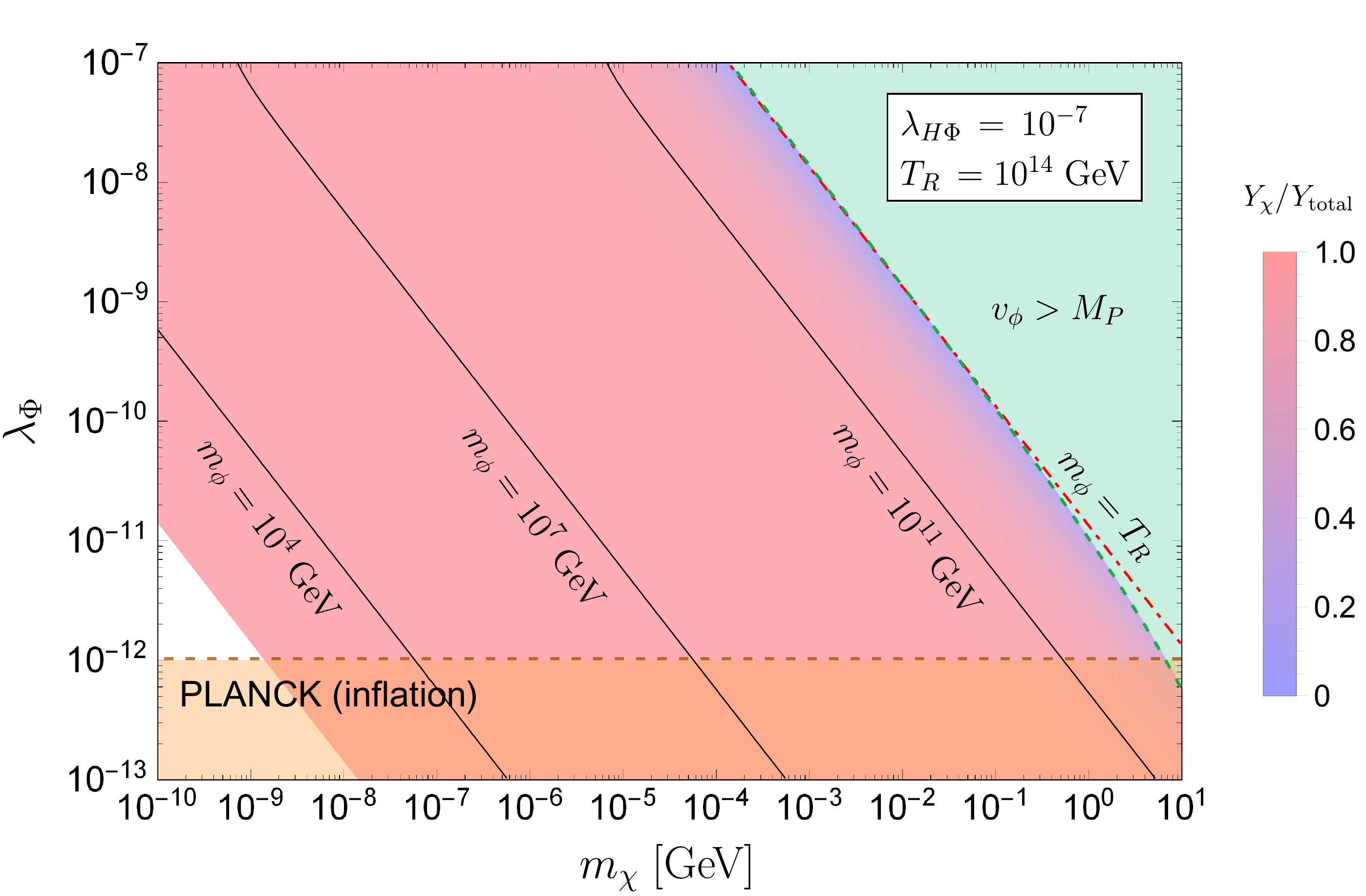}~
\includegraphics[scale=0.27]{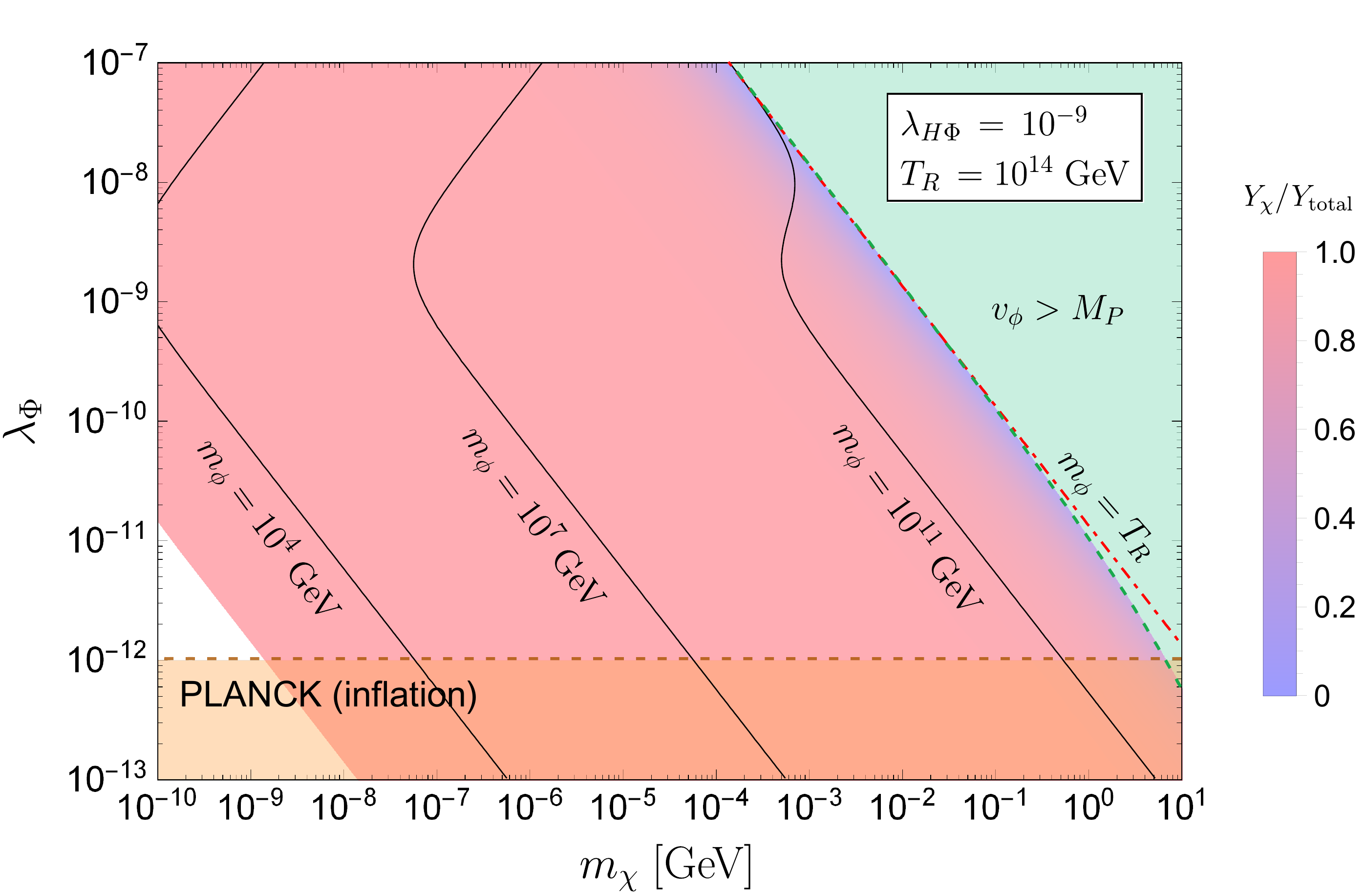}
\caption{
Same plots as Fig.~\ref{fig:allowedm_chi-lambda_Phi2} while 
the reheating temperature is taken to be a free parameter.
The portal coupling is chosen as
$\lambda_{H\Phi} =10^{-7}$ (left) and $10^{-9}$
(right), and the reheating temperature is 
$T_R=10^{10}~\mr{GeV}$ (above) and $10^{14}~\mr{GeV}$ (below).
\bigskip}
\label{fig:allowedm_chi-lambda_Phi3}
\end{figure}

In the above analysis, we have assumed that the reheating of the
universe simply occurs via the inflaton perturbative decay to the
Higgs field. However if there exists some other decay modes of the inflaton,
the reheating temperature generally takes a different value. In this situation,
the inflaton induced dark matter is estimated by
\begin{align}
 Y^{\mr{inf}}_\chi \approx
  \frac{3\sqrt{30}}{128\pi^2}\frac{g_*^{1/2}\lambda_\Phi M_P}{g_*^S\,T_R}
 \sqrt{1 - \frac{4 m_\chi^2}{m_\phi^2}},
\label{eq:direct2}
\end{align}
which is independent of the Higgs portal coupling $\lambda_{H\Phi}$
and almost independent of $m_\phi$ for a relatively heavy $\phi$.
The pNGB abundance from the UV freeze-in is also implicitly changed
since the parameter dependence of $T_R$ is modified from Eq.~\eqref{eq:tr}
to a free parameter, though the formula of $Y_D^\mr{UV}$ is still valid.
In the same fashion in Fig.~\ref{fig:allowedm_chi-lambda_Phi2}, we draw the 
parameter space realizing the correct dark matter abundance
via both of freeze-in and inflaton decay
(Fig.~\ref{fig:allowedm_chi-lambda_Phi3}). We find in almost all
allowed regions the freeze-in production is dominant. This comes from the fact
that the inflaton induced abundance Eq.~\eqref{eq:direct2} is almost
insensitive to $m_\phi$. Compared with
Fig.~\ref{fig:allowedm_chi-lambda_Phi2}, 
a heavier dark matter is possible up to $\mc{O}(10-100)~\mr{GeV}$
with a higher reheating temperature.

\section{Conclusion}
\label{sec:conclusion}

The pNGB dark matter model has been originally motivated from the fact that
the strong constraint of direct detection can naturally be evaded even
when it is a WIMP with sufficiently large couplings
with the SM particles. 
On the other hand, the pNGB dark matter can also be regarded as a
natural FIMP candidate if the VEV of the symmetry-breaking scalar is
large enough. That is because all couplings of the pNGB
are suppressed by the large VEV due to its NG property.

We have studied the model parameters for which the dark matter relic
is reproduced by the feeble couplings of pNGB, taking into account the
effect of thermal mass of the Higgs field.
The dark matter relic abundance is mainly determined by the mediator mass
and a smaller coupling of $\lambda_\Phi$ and $\lambda_{H\Phi}$ 
when the reheating temperature of the universe is larger than the
mediator mass. On the contrary, the abundance depends on 
the reheating temperature if it is not large as the mediator mass. 
These feature are similar to typical FIMPs.

We have also investigated the possibility that the radial component of
the symmetry-breaking scalar $\Phi$ plays a role of the inflaton.
Introducing the non-minimal coupling of $\Phi$ to gravity, 
the flat potential is understood by the rescaling,
and the parameter space consistent with the observations has been explored. 
We have found that the scalar self coupling is tiny
$\lambda_\Phi\gtrsim10^{-12}$ and the non-minimal coupling 
should be $\xi \gtrsim 10^{-2}$, which are rather different from the
Higgs inflation scenario. Furthermore, it is important   
the inflaton decay into the pNGB is unavoidable.
We have examined the allowed parameter regions taking into account
both of the freeze-in and the inflaton decay.
Combining these requirements for the dark matter relic and the
successful inflation, it is found that the pNGB FIMP dark matter
should be lighter than a few GeV when the freeze-in contribution is
assumed to be dominant. A heavier pNGB dark matter with 
$\mc{O}(10^{1-2})~\mr{GeV}$ mass is
possible if the inflaton-induced contribution comes to be effective
and/or the reheating process depends on some other dynamics.

In the present model, since the pNGB dark matter is stable due to the
$\mbb{Z}_2$ symmetry: $\chi \mapsto -\chi$  
coming from the CP invariance of the scalar potential, one may feel
there is no detectable signals from the pNGB dark matter with feeble couplings. 
However if the remnant symmetry is not exact as easily expected in a UV
completion of the model~\cite{Abe:2020iph}, the pNGB can decay into lighter
SM particles. 
Even if its couplings are highly suppressed by the large VEV, some
signals may be detectable in cosmic-ray observations.
That is left for future study.


\subsection*{Acknowledgments}
\noindent
The authors thank Tetsutaro Higaki and Takahiro Ohata for the useful
discussions and comments. 
The numerical computation in this work was carried out at the Yukawa
Institute Computer Facility.
This work is supported by JSPS Grant-in-Aid for Scientific Research KAKENHI
Grant No.~JP20J11901 (YA), JP20K22349 (TT), JP18H01214 and JP20K03949~(KY).

\appendix

\section{Thermal Mass Contribution}
\label{sec:Tmass}

Referring to \cite{Katz:2014bha}, we summarize the derivation of
thermal mass and its formula in the high temperature era.

We consider a field variable decomposed to the background
configuration $\sigma$ and its fluctuation $\rho$, 
and integrate out the latter.
Then the one-loop effective potential for the background $\sigma$ is given by
\begin{align}
 \mc{V}^T_{\mr{eff}}(\sigma) = \mc{V}_0(\sigma) + \mc{V}_1 (\sigma)+ \mc{V}^T_1 (\sigma;T),
\end{align}
where $\mc{V}_0(\sigma)$ is the classical potential for $\sigma$, and
$\mc{V}_1(\sigma)$, $\mc{V}^T_1(\sigma)$ are the one-loop contributions.
According to Refs.~\cite{Dolan:1973qd,Weinberg:1974hy},
the effective potential is evaluated on $\mbb{R}^3 \times S^1$ with the radius $1/T$
in order to take the thermal effect into account.

The effective potential for a real scalar is expressed as
\begin{align}
 \mc{V}_{1B}(\sigma) = \int \frac{d^3 k}{(2\pi)^3} \frac{E_{\bm{k}}}{2},
 \qquad
 \mc{V}^T_{1B}(\sigma;T) = \frac{T^4}{2\pi^2} J_B ( M /T) ,
\end{align}
where $E_{\bm{k}}^2 = \bm{k}^2 + M^2$, and $M$ denotes the $\rho$ mass 
in the $\sigma$ background. The mass $M$ is typically given by 
$
 M^2=M(\sigma)^2 = m_b^2 + \frac{\lambda}{2} \sigma^2
$
for the $\rho$ mass $m_b$ and the scalar self quartic coupling $\lambda$.
The function $J_B(y)$ is given by
\begin{align}
 J_B(y) = \int_0^\infty dx\, x^2 \, \log \Bigl( 1 - e^{- (x^2 + y^2)^{1/2}} \Bigr).
\end{align}
In the high-temperature region corresponding to $y \ll 1$,
the function $J_B(y)$ is approximately written as
\begin{align}
 J_B(y) \approx - \frac{\pi^4}{45} + \frac{\pi^2}{12} y^2 - \frac{\pi}{6} y^3 - \frac{y^4}{32} \log \biggl( \frac{y^2}{a_B} \biggr),
\end{align}
with $a_B =  \pi^2 e^{3/2 - 2 \gamma_E}$ and $\gamma_E$ is the Euler constant.
The $y^2$ term contributes to the thermal mass.

For a Dirac fermion, the one-loop effective potential is given by
\begin{align}
 \mc{V}_{1F}(\sigma) = -4 \int \frac{d^3k}{(2\pi)^3} \frac{E_{\bm{k}}}{2} ,
 \qquad
 \mc{V}_{1F}^T (\sigma ;T) = - \frac{2T^4}{\pi^2} J_F(M/T),
\end{align}
with
\begin{align}
 J_F(y) = - \int_0^\infty dx\, x^2 \log \biggl( 1 + e^{- (x^2+ y^2)^{1/2}} \biggr).
\end{align}
The mass in the $\sigma$ background is typically given by using a
Yukawa coupling $g$ as
$
 M=M(\sigma) = m_f + g \frac{\sigma}{\sqrt{2}},
$
for the fermion mass $m_f$. As in the bosonic case,
the function $J_F(y)$ is approximately written as
\begin{align}
 J_F(y) \approx \frac{7 \pi^4}{360} - \frac{\pi^2}{24} y^2 - \frac{y^4}{32} \log \biggl( \frac{y^4}{a_F} \biggr),
\end{align}
with $a_F =  16 \pi^2 e^{3/2 - 2 \gamma_E}$
in the high-temperature region corresponding to $y \ll 1$.

The thermal mass is defined from the one-loop effective potential
$\mc{V}^T_1(\sigma ;T)$ as
\begin{align}
 \Delta \equiv \frac{\partial^2 \mc{V}^T_1}{\partial\sigma^2}\biggr|_{\sigma=0}.
\end{align}
Using the above formulae, the thermal mass contributions of particle
$i$ are given by
\begin{align}
 \left\{\begin{array}{lcl}
 \Delta^B = g_i \frac{T^4}{2 \pi^2} \frac{\pi^2}{12} \frac{M^2(\sigma)''}{T^2}
 = \frac{g_i T^2 M^2 (\sigma)''}{24} && \text{for bosons}
 \\[2mm]
 \Delta^F = - g_i \frac{T^4}{2\pi^2} \bigl( \frac{- \pi^2}{24} \bigr) \frac{M^2(\sigma)''}{T^2}
 = \frac{g_i T^2 M^2(\sigma)''}{48} && \text{for fermions}
\end{array}
\right.
\end{align}
where $g_i$ is the degrees of freedom of the particle $i$ and
the prime means the derivative with respect to $\sigma$.

For the SM Higgs boson, there are three sources of thermal mass;
electroweak gauge bosons, quarks and leptons, and Higgs
scalar. Applying the above result to the SM, the thermal mass
contributions are found
\begin{align}
 \Delta^{\mr{gauge}}_h ~=&~ \frac{g_Y^2}{16} T^2 + \frac{3 g_2^2}{16} T^2,
 \\
  \Delta^{\mr{fermion}}_h =&~ \frac{T^2}{12} \Bigl[
 y_e^2 + y_\mu^2 + y_\tau^2 + 3 \bigl( y_u^2 + y_d^2 + y_c^2 + y_s^2 + y_t^2 + y_b^2 \bigr)
 \Bigr]
 \approx \frac{y_t^2}{4} T^2,
 \\
  \Delta^{\mr{scalar}}_h 
~=&~ \frac{\lambda_H}{4} T^2.
\end{align}


\end{document}